\documentclass[a4paper,12pt]{iopart}
\pdfoutput=1
\usepackage[utf8]{inputenc}
\usepackage{iopams} % For ams fonts and symbols in iopart.cls
\usepackage{dsfont} % For double-stroke 1 as $\mathds{1}$ since $\mathbb{1}$ doesn't exist in all fonts
\usepackage[british]{babel}
\usepackage{microtype} % Microtypography!
\usepackage{xspace}
\usepackage[bookmarksnumbered=true]{hyperref}

\eqnobysec % Label equations in sections

\newcommand{\D}{\mathrm{d}} % Differential d
\newcommand{\I}{\mathrm{i}} % Imaginary unit

\newcommand{\eqref}[1]{\eref{#1}}

\newcommand{\KGe}{Klein--Gordon equation\xspace}

\begin{document}

\title{Post-Newtonian corrections to Schrödinger equations in gravitational fields}
\author{Philip K Schwartz$^1$ and Domenico Giulini$^{1,2}$}
\address{$^1$ Institute for Theoretical Physics, Leibniz University Hannover, Appelstraße 2, 30167 Hannover, Germany}
\address{$^2$ Center of Applied Space Technology and Microgravity, University of Bremen, Am Fallturm 1, 28359 Bremen, Germany}
\eads{\mailto{philip.schwartz@itp.uni-hannover.de}, \mailto{giulini@itp.uni-hannover.de}}

\begin{abstract}
	In this paper we extend the WKB-like `non-relativistic' expansion of the minimally coupled \KGe after Kiefer and Singh \cite{kiefer91}, Lämmerzahl \cite{laemmerzahl95} and Giulini and Großardt \cite{giulini12} to arbitrary order in $c^{-1}$, leading to Schrödinger equations describing a quantum particle in a general gravitational field, and compare the results with canonical quantisation of a free particle in curved spacetime, following Wajima \emph{et al.} \cite{wajima97}.
	Furthermore, using a more operator-algebraic approach, the \KGe and the canonical quantisation method are shown to lead to the same results for some special terms in the Hamiltonian describing a single particle in a general stationary spacetime, without any `non-relativistic' expansion.
\end{abstract}


%%%%%%%%%%%%%%%%%%%%%%%%%%%%%%%%%%%%%%%%%%%%%%%%%%%%%%%%%%%
\section{Introduction} \label{sec:Introduction}

Suppose we are given a quantum-mechanical system whose time evolution in the absence of gravity is known in terms of the ordinary time-dependent Schrödinger equation. In other words: We know the system's Hamiltonian if all gravitational interactions are neglected.
We ask: what principles do we use in order to deduce the system's interaction with a given external gravitational field?
Note that by `gravitational field' we understand all the ten components $g_{\mu\nu}$ of the spacetime metric that are subject to Einstein's field equations of general relativity (or, more generally, to the equations of some other metric theory of gravity), not just the scalar component $\phi$ representing the Newtonian potential.

The behaviour of quantum systems in general gravitational fields is naturally of fundamental interest, relating, e.g., to the reaction of quantum systems to gravitational waves \cite{gao18}, tests of general relativity, the controversially discussed topic of gravitationally induced quantum dephasing \cite{pikovski15,bonder15,pang16}, quantum tests of the classical equivalence principle \cite{schlippert14} or proposals of quantum formulations of the equivalence principle \cite{zych18} and tests thereof \cite{rosi17}.

The conceptual difficulty we are addressing here has to do with the fact that the usual `minimal coupling scheme' that we usually employ in order to couple any classical field obeying Lorentz invariant equations of motion simply does not apply to the case at hand.
We recall that, in a nutshell, this prescription is a two step process:
1)~Write down the matter's dynamical law in a Poincaré invariant fashion in Minkowski spacetime;
2)~replace the flat Minkowski metric $\eta$ by a general Lorentzian metric $g$ and the partial derivatives with respect to the affine coordinates of Minkowski spacetime (i.e. the covariant derivatives with respect to $\eta$) by the Levi--Civita covariant derivative with respect to $g$.

This procedure suffers from an essential non-uniqueness which has to do with the fact that the covariant derivatives with respect to $\eta$ commute whereas those with respect to $g$ generally do not. Hence there are equivalent forms for the dynamical laws in step~1) that turn into inequivalent extensions after performing step~2).
The differences will consist in terms that result from commuting covariant derivatives and hence in local couplings of $g$'s curvature tensor to the matter field. Modulo this well known ambiguity the prescription just outlined is straightforward to apply and has been successfully used in all applications of general relativity.

The point addressed here in connection with quantum mechanics is not so much the ambiguity just described, but rather the obvious failure to even implement step~1): There simply is no obvious way to rewrite the Schrödinger equation into a Poincaré invariant form in order to be able to apply step~2).
Special relativity turns quantum mechanics into relativistic quantum field theory (RQFT) in which the particle concept ceases to be meaningful in presence of general background geometries~\cite{baer09,wald94}. Does that mean we would have to employ the whole machinery of RQFT in order to just answer simple questions concerning matter-gravity interactions that go beyond the simplest (and obvious; see below) couplings to the Newtonian potential?
We think the answer is no, but at the same time we think that the alternative should not result in \emph{ad hoc} procedures guided by more or less well founded `physical intuition'. Rather we should look for a general and systematic method that allows to derive the full coupling by means of an algorithm that arguably qualifies as a proper post-Newtonian approximation. We see this paper as a positive contribution to this end.

To be slightly more precise, we recall that the Schrödinger equation describing a `non-relativistic' (i.e. not Poincaré invariant)
%\footnote{Even if the common terminology is
%not quite correct, we will follow it here and
%mean `Poincaré-relativistic' by `relativistic'
%and use the phrase `non-relativistic' to refer
%to Galilei-relativistic systems, or to
%approximation of Poincaré symmetry by Galilei
%symmetry etc.}
particle of mass $m$ and zero spin in a background Newtonian gravitational field with potential $\phi$ is
\begin{equation}
%\label{eq:SchroedingerEquation}
\I\hbar\partial_t \psi = \left(-\frac{\hbar^2}{2m}\Delta + m \phi\right) \psi\,.
\end{equation}
This equation has extensively been tested in the gravitational field of the earth, beginning with neutron interferometry in the classic COW experiment \cite{COW75} and leading up to atom interferometers of the Kasevich--Chu type, accomplishing, e.g., highly precise measurements of the gravitational acceleration $g$ on the earth \cite{farah14}.
We now ask what kind of `post-Newtonian corrections' to this equation arise from general relativity or other metric theories of gravity, considering additional terms involving the Newtonian potential $\phi$ as well as new terms involving all ten metric components $g_{\mu\nu}$.

In the existing literature, one finds two different main approaches to this problem of post-Newtonian correction terms for the Schrödinger equation describing a particle in a curved spacetime. The first, described, e.g., by Wajima \emph{et al.} \cite{wajima97}, starts from a classical description of the particle and applies canonical quantisation rules adapted to the situation (in a somewhat \emph{ad hoc} fashion) to derive a quantum mechanical Hamiltonian. By an expansion in powers of $c^{-1}$ (at the stage of the classical Hamiltonian), one finds the desired correction terms.
Other methods along a similar line use path integral quantisation on the classical system, as, e.g., the semi-classical calculation by Dimopoulos \emph{et al.} \cite{dimopoulos08}.

The second, fundamentally different approach takes a field-theoretic perspective and derives the Schrödinger equation as an equation for the positive frequency solutions of the minimally coupled classical \KGe\footnote
	{For simplicity we call, as is common, the Poincaré invariant mass $m$ and zero-spin equation by the names of Oskar Klein and Walter Gordon, even though Erwin Schrödinger and Vladimir Fock were amongst the first to consider it.}.
This is accomplished by Kiefer and Singh \cite{kiefer91}, Lämmerzahl \cite{laemmerzahl95} and Giulini and Großardt \cite{giulini12} by making a WKB-like ansatz for the Klein--Gordon field and thus (formally) expanding the \KGe in powers of $c^{-1}$, in the end viewing the Klein--Gordon theory as a (formal) deformation of the Schrödinger theory, implementing the deformation of Galilei to Poincaré symmetry well-known at the level of Lie algebras \cite{inonu53}. This second method seems to be more firmly rooted in first principles than the canonical quantisation method, since it can at least heuristically be motivated from quantum field theory in curved spacetimes (see section~\ref{subsec:WKB_heur}).

The calculation of general-relativistic corrections to phase shifts in atom interferometry by Dimopoulos \emph{et al.} \cite{dimopoulos08} used a \emph{semi-classical} approximation scheme, the so-called `stationary phase approximation' in path integral terms, which is non-exact for Hamiltonians of higher than quadratic order in positions and momenta. The approximation breaks down for wave packets with large width in phase space, so that similar methods cannot directly be applied when considering the effect of general gravitational fields on, e.g., large spatial superpositions, for example proposed relevant for gravitational dephasing by Pikovski \emph{et al.} \cite{pikovski15}. In contrast to this, using a modified Schrödinger equation makes it possible to, in principle, describe quantum systems under gravity without any semi-classical approximation.

Although the two methods for obtaining post-Newtonian Schrödinger equations described above are very different in spirit, they lead to comparable results in lowest orders. To make possible a general comparison beyond the explicit examples considered in the existing literature\footnote
	{Wajima \emph{et al.} \cite{wajima97} considered a first-order post-Newtonian metric for a point-like rotating source, Lämmerzahl \cite{laemmerzahl95} used a first-order Eddington--Robertson PPN metric.},
we will apply the methods to as general a metric as possible: In section \ref{sec:canon_quant}, we will give a brief overview over the canonical quantisation method, while section \ref{sec:WKB} will develop the WKB-like formal expansion of the \KGe to arbitrary order in $c^{-1}$ in a general metric given as a formal power series in $c^{-1}$, significantly extending existing explicit examples to the general case. This leads to some simple comparisons of the resulting Hamiltonian with the one from canonical quantisation.

In section \ref{sec:mom_exp}, we consider a (formal) expansion of the \KGe in powers of momentum operators leading to a Schrödinger form of the equation, yielding some general statements about similarities between the canonical and the Klein--Gordon methods without any `non-relativistic' expansion in powers of $c^{-1}$.

A general WKB-like `non-relativistic' formal expansion of the \KGe to obtain a Schrödinger equation was already considered by Tagirov in \cite{tagirov90} and a series of follow-up papers \cite{tagirov92,tagirov96}, as summarised in \cite{tagirov99}; but unlike our approach, these works did not expand the metric, thus not allowing to directly apply the results to metrics given as a power series in $c^{-1}$. Tagirov also compared his WKB-like approach to methods of canonical quantisation \cite{tagirov03}, but did this only for the case of static metrics.

We will use the `mostly plus' signature convention for the spacetime metric, i.e. $(-,+,+,+)$. Since we are concerned mostly with conceptual questions, we will generally not be mathematically very rigorous, and in particular not mention domains of definition of operators.

%%%%%%%%%%%%%%%%%%%%%%%%%%%%%%%%%%%%%%%%%%%%%%%%%%%%%%%%%%%
\section{Canonical quantisation of a free particle} \label{sec:canon_quant}

In the following, we will describe the canonical quantisation approach used by Wajima \emph{et al.} \cite{wajima97} to derive a Hamiltonian for a quantum particle in the post-Newtonian gravitational field of a point-like rotating source for the analysis of effects on interferometric phase shifts. We will focus on the conceptual issues of the procedure since we want to stay as general as possible.

The classical action for a `relativistic' point particle of mass $m$ in curved spacetime with metric $g$ is
\begin{equation} \label{eq:class_action}
	S = -mc \int \sqrt{-g(x'(\lambda),x'(\lambda))} \D \lambda = -mc \int \sqrt{-g_{\mu\nu}x'^\mu x'^\nu} \D \lambda,
\end{equation}
where $x(\lambda)$ is the arbitrarily parametrised worldline of the particle.

We assume the spacetime to be globally hyperbolic and perform a $3+1$ decomposition of spacetime \cite{giulini14}: We foliate spacetime $M$ into 3-dimensional spacelike Cauchy surfaces $\Sigma_t$ which are images of an `abstract' Cauchy surface $\Sigma$ under a family of embeddings $\mathcal E_t \colon \Sigma \to M$, parametrised by a `foliation parameter' $t \in \mathbb R$, and we introduce spacetime coordinates such that $x^i$ are coordinates on $\Sigma$ and $x^0 = ct$. Parametrising the worldline of the particle by $t$, the classical Hamiltonian can be computed to be
\begin{equation} \label{eq:class_Ham}
	H = \frac{1}{\sqrt{-g^{00}}} c \left[m^2 c^2 + \left(g^{ij} - \frac{1}{g^{00}} g^{0i} g^{0j}\right) p_i p_j\right]^{1/2} + \frac{c}{g^{00}} g^{0i} p_i,
\end{equation}
when expressed in terms of the spacetime metric,
where $p_i$ are the momenta conjugate to $x^i$. Full details of this calculation can be found in \ref{app:details_class_Ham}.

Now, we want to `canonically quantise' this Hamiltonian. To this end, we have to define a Hilbert space for our quantum theory, define position and momentum operators acting on this Hilbert space and satisfying the canonical commutation relations, and choose an operator ordering scheme for symmetrising products of momenta and (functions of) position. By then expanding the square root in the classical Hamiltonian \eqref{eq:class_Ham} to the desired order in $c^{-1}$ (or in momenta, see section \ref{sec:mom_exp}) and afterwards replacing the classical momentum and position variables by the corresponding operators, applying the chosen ordering scheme, we get a quantised Hamiltonian $\hat H$ acting on our Hilbert space and can postulate a Schrödinger equation in the usual form
\begin{equation}
	\I\hbar\partial_t \psi = \hat H \psi.
\end{equation}
Let us stress once more that this Hamiltonian will depend not only on the choice of $3+1$ decomposition of spacetime into Cauchy surfaces, which is the case since for a post-Newtonian approximation we need some notion of separate space and time, but also on the choice of operator ordering scheme, which we leave open in order to keep the discussion as general as possible.

We now turn to the definition of the Hilbert space and the position and momentum operators, which is more subtle than might be anticipated at first thought. Since the position variables in the classical Hamiltonian \eqref{eq:class_Ham} are the spatial coordinates $x^i$ on the Cauchy surfaces, we want the quantum position operators to directly correspond to these. Thus, we want to define the Hilbert space as some space of square-integrable `wavefunctions' of the $x^i$, such that we can take as position operators simply the operators of multiplication with the coordinates. The question of definition of the Hilbert space thus becomes a question of choice of a scalar product on (some subspace of) the space of functions of the $x^i$.

To be more precise, we do not just need a single Hilbert space: To any `time' (i.e. foliation parameter) $t$ we want to associate a wavefunction $\psi(t)$ giving rise to a position probability distribution \emph{on the Cauchy surface $\Sigma_t$ corresponding to $t$}, so we need to consider an individual Hilbert space for each Cauchy surface in the $3+1$ decomposition. But since we want to relate these wavefunctions by a Schrödinger equation, we have to somehow identify the Hilbert spaces corresponding to the different Cauchy surfaces, which we are now going to explain.

A natural, geometric choice of scalar product on the space of functions on $\Sigma_t$ is the $\mathrm{L}^2$-scalar product with respect to the induced metric measure (cf. \cite{wajima97}), i.e.
\begin{equation} \label{eq:canon_scalar_prod_induced}
	\langle \psi, \varphi \rangle_{\Sigma_t} := \int \overline\psi \varphi \sqrt{^{(3)}g|_{\Sigma_t}} \D^3x,
\end{equation}
where here and in the following, we write ${^{(3)}g} = \det(g_{ij})$. Consider first the case that the spatial metric $g_{ij}$ be independent of $t$, i.e. that the induced geometry be the same for all Cauchy surfaces. Then this scalar product is independent of $t$, such that the Hilbert spaces corresponding to the different Cauchy surfaces are canonically identified by simply identifying the wavefunctions\footnote{Implicitly identifying functions defined on $\Sigma_t = \mathcal E_t(\Sigma)$ with functions on $\Sigma$ via $\mathcal E_t$, thus using once more the data of the $3+1$ decomposition.}. We can then define the momentum operator as
\begin{equation} \label{eq:canon_mom_op_induced}
	\hat p_i := -\I\hbar {^{(3)}g^{-1/4}}\partial_i ({^{(3)}g^{1/4}}\cdot),
\end{equation}
which is symmetric with respect to the scalar product and fulfils the canonical commutation relation $[x^i,\hat p_j] = \I \hbar \delta^i_j$, and carry out canonical quantisation as described above.

If we allow for $g_{ij}$ to depend on $t$, the scalar product \eqref{eq:canon_scalar_prod_induced} depends on $t$ and thus the canonical map $\mathrm{L}^2(\Sigma, \langle\cdot, \cdot\rangle_{\Sigma_t}) \ni \psi \mapsto \psi \in \mathrm{L}^2(\Sigma, \langle\cdot, \cdot\rangle_{\Sigma_s})$ no longer is an isomorphism of Hilbert spaces. I.e. the natural identification from above does not work, spoiling the program of canonical quantisation. A natural solution to this problem is to instead consider the time-independent `flat' $\mathrm L^2$-scalar product
\begin{equation}
	\langle \psi_f, \varphi_f \rangle_f := \int \overline{\psi_f} \varphi_f \D^3x
\end{equation}
together with the `flat' momentum operator $\bar p_i := -\I\hbar \partial_i$. Using these, canonical quantisation can be carried out. At first sight, this scalar product could seem less `geometric' than \eqref{eq:canon_scalar_prod_induced}, but it can be seen to have as much invariant meaning as the latter by realising that, geometrically speaking, the `flat' wavefunctions $\psi_f, \varphi_f$ be scalar \emph{densities} (of weight $1/2$) instead of scalar functions. Since this choice of `flat' scalar product can be applied to more general situations, and it eases the comparison to usual non-relativistic Schrödinger theory and to the WKB method, we will adopt it from now on, i.e. `canonically quantise' the expanded classical Hamiltonian by replacing the classical momentum by the flat momentum operator (applying our chosen ordering scheme).

In fact, in the case of time-independent $g_{ij}$ the quantum theories resulting from the two choices described above are unitarily equivalent by the Stone--von Neumann theorem, via the unitary operator given by $\psi \mapsto \psi_f = {^{(3)}g^{1/4}} \psi$.

%%%%%%%%%%%%%%%%%%%%%%%%%%%%%%%%%%%%%%%%%%%%%%%%%%%%%%%%%%%
\section{WKB-like expansion of the \KGe} \label{sec:WKB}

Now, we will consider WKB-like formal expansions in $c^{-1}$ of the classical, minimally coupled \KGe for a particle of mass $m > 0$,
\begin{equation} \label{eq:KG}
	\left(\Box - \frac{m^2 c^2}{\hbar^2}\right)\Psi_\mathrm{KG} = 0,
\end{equation}
leading to a Schrödinger equation with post-Newtonian corrections, as considered by Lämmerzahl in \cite{laemmerzahl95} for a simple PPN metric in Eddington--Robertson parametrisation and by Giulini and Großardt in \cite{giulini12} for general spherically symmetric metrics.

Instead of \eqref{eq:KG} one could also consider the more general case of a possibly non-minimally coupled \KGe, i.e. including some curvature term. This is customary in modern literature on quantum field theory in curved spacetime, where an additional term $-\xi R \Psi_\mathrm{KG}$ is included in the equation, $R$ being the scalar curvature of the spacetime \cite[eq. (5.57)]{baer09}. In particular, for the choice of $\xi = \frac{1}{6}$ (`conformal coupling'), the equation becomes conformally invariant in the massless case $m=0$, and also in the massive case there are some arguments favouring the conformally coupled \KGe, in particular in de Sitter spacetime \cite{tagirov73}. Nevertheless, we will for the sake of simplicity stick with the minimally coupled equation in this paper, leaving non-minimal coupling for later, possibly more coordinate-independent investigations.

After giving a heuristic motivation for consideration of the classical \KGe, we will describe the expansion of the \KGe to arbitrary order in $c^{-1}$, then explain the transformation to a `flat' $\mathrm L^2$-scalar product for comparison to canonical quantisation and finally consider the metric of the Eddington--Robertson PPN test theory as a simple explicit example.

%%%%%%%%%%%%%%%%%%%%%%%%%%%%%%%%%%%%%%%%%%%%%%%%%%%%%%%%%%%
\subsection{Heuristic motivation from quantum field theory in stationary spacetimes} \label{subsec:WKB_heur}

As already mentioned in the introduction, consideration of the classical \KGe can, on a heuristic level, be motivated from quantum field theory in curved spacetimes. Namely, the quantum field theory construction for the free Klein--Gordon field on a globally hyperbolic stationary spacetime proceeds as follows \cite[section 4.3]{wald94}.

We consider the \KGe \eqref{eq:KG} and the Klein--Gordon inner product, which for two solutions $\Psi_\mathrm{KG}, \Phi_\mathrm{KG}$ of \eqref{eq:KG} is given by
\begin{eqnarray}\eqalign{ \label{eq:KG_ip_general}
	\langle\Psi_\mathrm{KG},\Phi_\mathrm{KG}\rangle_\mathrm{KG} &= \I\hbar c \int_\Sigma n^\nu [\overline{\Psi_\mathrm{KG}} (\nabla_\nu \Phi_\mathrm{KG}) - (\nabla_\nu \overline{\Psi_\mathrm{KG}}) \Phi_\mathrm{KG}] \sqrt{^{(3)}g} \D^3x\\
	&= \I\hbar c \int_\Sigma n^\nu [\overline{\Psi_\mathrm{KG}} (\partial_\nu \Phi_\mathrm{KG}) - (\partial_\nu \overline{\Psi_\mathrm{KG}}) \Phi_\mathrm{KG}] \sqrt{^{(3)}g} \D^3x
}\end{eqnarray}
where $\Sigma$ is a spacelike Cauchy surface, $^{(3)}g$ is the determinant of the induced metric on $\Sigma$ and $n$ is the future-directed unit normal vector field of $\Sigma$. In the second line, which is valid in a coordinate basis, we used that the covariant derivative of a scalar function is just the ordinary derivative, i.e. given by a partial derivative in the case of a coordinate basis. Using the \KGe and Gauß' theorem, \eqref{eq:KG_ip_general} can be shown to be independent of the choice of $\Sigma$ under the assumption that the fields satisfy suitable boundary conditions.

The Hilbert space of the quantum field theory is now the bosonic Fock space over the `one-particle' Hilbert space constructed, loosely speaking, as the completion of the space of classical solutions of the \KGe with `positive frequency' (wrt. the stationarity Killing field) with the Klein--Gordon inner product.

To be more precise, the construction of the `one-particle' Hilbert space is a little more involved, since it is not \emph{a priori} clear what is meant by `positive frequency solutions': At first, the space of classical solutions of the \KGe is completed in a certain inner product to obtain an `intermediate' Hilbert space on which the generator of time translations (wrt. the stationarity Killing field) can be shown to be a self-adjoint operator; the positive spectral subspace of this operator is then completed in the Klein--Gordon inner product to give the Hilbert space of one-particle states. For details on the construction, see \cite[section 4.3]{wald94} and the references cited therein.

So the one-particle sector of the free Klein--Gordon quantum field theory in globally hyperbolic stationary spacetime is described by an appropriate notion of positive frequency solutions of the classical \KGe, using the Klein--Gordon inner product.

At this point, the quantum field theoretic motivation of the WKB-like method becomes merely heuristic: Since in the following we will not solve the \KGe exactly, but consider formal expansions of it in powers of $c^{-1}$, it will not be possible to exactly determine the space of positive frequency solutions according to the procedure described above; instead, we will merely choose an oscillating phase factor such as to guarantee the solution to have positive instead of negative frequency in lowest order in the expansion (see \eqref{eq:WKB_pos_freq}). If analysed more rigorously, it could turn out that for an asymptotic solution to be of positive frequency in some stricter sense, additional restrictions on the solution have to be made, possibly altering the function space under consideration. I.e. in principle, this could lead to the Hamiltonian we will obtain being altered when considering a rigorous post-Newtonian expansion of quantum field theory in curved spacetime.

In the non-stationary case, there is no canonical notion of particles and thus, strictly speaking, the whole question about the behaviour of single quantum particles does not make sense. Nevertheless, for an observer moving on an orbit which is approximately Killing, the classical Klein--Gordon theory can, on a heuristic level, still be expected to lead to approximately correct predictions regarding this observer's observations.

Even if this motivation is just a heuristic, the WKB-like approach of expanding the \KGe in powers of $c^{-1}$ allows us to view the classical Klein--Gordon theory as a formal deformation of the `non-relativistic' Schrödinger theory, and makes the sense in which that happens formally precise.

%%%%%%%%%%%%%%%%%%%%%%%%%%%%%%%%%%%%%%%%%%%%%%%%%%%%%%%%%%%
\subsection{General derivation}

We fix a coordinate system $(x^0 = ct, x^i)$ and assume that the components of the inverse metric be (formal) power series
\begin{equation} \label{eq:exp_metric}
	g^{\mu\nu} = \eta^{\mu\nu} + \sum_{k=1}^\infty c^{-k} g^{\mu\nu}_{(k)}
\end{equation}
in the inverse of the velocity of light $c^{-1}$, the lowest-order term being given by the (inverse) Minkowski metric. To perform the `non-relativistic' expansion in the correct way, we have to explicitly include $c^{-1}$ in the definition of coordinate time $t = c^{-1} x^0$.

In coordinates, the d'Alembert operator in a general Lorentz metric is given by
\begin{eqnarray}\eqalign{ \label{eq:box_op}
	\Box f &= \frac{1}{\sqrt{-g}} \partial_\mu(\sqrt{-g} g^{\mu\nu} \partial_\nu f)\\
	&= \frac{1}{\sqrt{-g}} (\partial_\mu\sqrt{-g}) g^{\mu\nu} \partial_\nu f + \partial_\mu(g^{\mu\nu}) \partial_\nu f + g^{\mu\nu} \partial_\mu\partial_\nu f,
}\end{eqnarray}
where $g = \det(g_{\mu\nu})$. The second and third term in this expression can easily be expanded in $c^{-1}$ by inserting the expansion \eqref{eq:exp_metric} of the components of the inverse metric.

Since the remaining first term involves the expression
\begin{equation}
	\frac{1}{\sqrt{-g}} \partial_\mu\sqrt{-g} = \frac{1}{2g} \partial_\mu g = \frac{1}{2} g^{\rho\sigma} \partial_\mu g_{\rho\sigma} = -\frac{1}{2} g_{\rho\sigma} \partial_\mu g^{\rho\sigma},
\end{equation}
we need an expression for the $c^{-1}$-expansion of the components of the \emph{metric}.
Rewriting the expansion of the inverse metric as
\begin{equation}
	g^{\mu\nu} = \left[ \left(\mathds{1} + \sum_{k=1}^\infty c^{-k} g^{-1}_{(k)} \eta\right) \eta^{-1} \right]^{\mu\nu},
\end{equation}
we see that a formal Neumann series can be used to invert the power series. Using the Cauchy product formula, the first term of \eqref{eq:box_op} can then be explicitly expanded. Full details of this calculation can be found in \ref{app:details_WKB}.

Combining the results for the three terms, the full expansion of the d'Alembert operator reads
\begin{eqnarray}\fl\eqalign{ \label{eq:box_expanded}
	\Box f &= \frac{1}{2} \sum_{k=4}^\infty c^{-k} \sum_{l+m = k-2} \sum_{n=1}^\infty (-1)^n \frac{1}{n} g^{00}_{(m)} [\partial_t \tr(\eta g^{-1}_{(l,n)})] \partial_t f \\
	&\qquad + \frac{1}{2} \sum_{k=3}^\infty c^{-k} \sum_{l+m = k-1} \sum_{n=1}^\infty (-1)^n \frac{1}{n} g^{0i}_{(m)} \left([\partial_t \tr(\eta g^{-1}_{(l,n)})] \partial_i f + [\partial_i \tr(\eta g^{-1}_{(l,n)})] \partial_t f\right)\\
	&\qquad - \frac{1}{2} \sum_{k=3}^\infty c^{-k} \sum_{n=1}^\infty (-1)^n \frac{1}{n} [\partial_t \tr(\eta g^{-1}_{(k-2,n)})] \partial_t f\\
	&\qquad + \sum_{k=3}^\infty c^{-k} (\partial_t g^{00}_{(k-2)}) \partial_t f + \sum_{k=3}^\infty c^{-k} g^{00}_{(k-2)} \partial_t^2 f\\
	&\qquad + \frac{1}{2} \sum_{k=2}^\infty c^{-k} \sum_{l+m = k} \sum_{n=1}^\infty (-1)^n \frac{1}{n} g^{ij}_{(m)} [\partial_i \tr(\eta g^{-1}_{(l,n)})] \partial_j f\\
	&\qquad + \sum_{k=2}^\infty c^{-k} \left((\partial_t g^{0i}_{(k-1)}) \partial_i f + (\partial_i g^{0i}_{(k-1)}) \partial_t f\right) + \sum_{k=2}^\infty c^{-k} 2 g^{0i}_{(k-1)} \partial_t\partial_i f - c^{-2}\partial_t^2 f\\
	&\qquad + \frac{1}{2} \sum_{k=1}^\infty c^{-k} \sum_{n=1}^\infty (-1)^n \frac{1}{n} [\partial_i \tr(\eta g^{-1}_{(k,n)})] \partial_i f\\
	&\qquad + \sum_{k=1}^\infty c^{-k} (\partial_i g^{ij}_{(k)}) \partial_j f + \sum_{k=1}^\infty c^{-k} g^{ij}_{(k)} \partial_i\partial_j f + \Delta f,
}\end{eqnarray}
where latin indices are `spatial' indices running from 1 to 3,
$\Delta = \sum_{i=1}^3 \partial_i^2$ denotes the `flat' Euclidean Laplacian in the spatial coordinates,
in sums like $\sum_{l+m = k}$, the indices $l$ and $m$ take values $\ge 1$,
and we introduced the notation
\begin{equation}
	g^{-1}_{(k,n)} := \sum_{i_1+\cdots+i_n = k \atop 1 \le i_1,\ldots,i_n \le k} g^{-1}_{(i_1)}\eta g^{-1}_{(i_2)}\eta \cdots g^{-1}_{(i_n)}.
\end{equation}

Now, we make the WKB-like ansatz
\begin{equation} \label{eq:WKB_ansatz}
	\Psi_\mathrm{KG} = \exp\left(\frac{\I c^2}{\hbar}S\right) \psi, \psi = \sum_{k=0}^\infty c^{-k} a_k
\end{equation}
for the Klein--Gordon field (cf. \cite{giulini12}), where $S$ is a \emph{real} function; i.e. we separate off a phase factor and expand the remainder as a power series in $c^{-1}$. All the functions $S, a_k$ are assumed to be independent of the expansion parameter $c^{-1}$. The derivatives of the field are
\begin{equation}
	\partial_\mu \Psi_\mathrm{KG} = \frac{\I c^2}{\hbar} (\partial_\mu S) \Psi_\mathrm{KG} + \exp(\ldots) \partial_\mu\psi
\end{equation}
and
\begin{eqnarray} \eqalign{
	\partial_\mu\partial_\nu \Psi_\mathrm{KG} &= \exp\left(\frac{\I c^2}{\hbar}S\right) \bigg({- \frac{c^4}{\hbar^2}} (\partial_\mu S)(\partial_\nu S) \psi + \frac{\I c^2}{\hbar} \big[(\partial_\mu\partial_\nu S) \psi\\
		&\qquad+ (\partial_\mu S) \partial_\nu\psi + (\partial_\nu S) \partial_\mu\psi\big] + \partial_\mu\partial_\nu \psi\bigg).
}\end{eqnarray}

Using this and the expansion \eqref{eq:box_expanded} of the d'Alembert operator, we can now analyse the \KGe \eqref{eq:KG} order by order in $c^{-1}$. At the lowest occurring order $c^4$, we get
\begin{equation}
	-\exp\left(\frac{\I c^2}{\hbar}S\right) \frac{1}{\hbar^2} (\partial_i S)(\partial_i S) a_0 = 0,
\end{equation}
which is equivalent\footnote{For nontrivial solutions, i.e. $a_0 \ne 0$.} to $\partial_i S = 0$.
So $S$ is a function of (coordinate) time only. Using this, the \KGe has no term of order $c^3$.

At $c^2$, we get
\begin{equation}
	\exp\left(\frac{\I c^2}{\hbar}S\right) \left(\frac{1}{\hbar^2} (\partial_t S)^2 - \frac{m^2}{\hbar^2}\right) a_0 = 0,
\end{equation}
equivalent to $\partial_t S = \pm m$. Since we are interested in positive-frequency solutions of the \KGe, we choose $\partial_t S = -m$, leading to
\begin{equation} \label{eq:WKB_pos_freq}
	S = -mt
\end{equation}
(an additional constant term would lead to an irrelevant global phase).

The $c^1$ coefficient leads to the equation
\begin{equation}
	-\exp\left(\frac{\I c^2}{\hbar}S\right) g^{00}_{(1)} \frac{m^2}{\hbar^2} a_0 = 0,
\end{equation}
equivalent to
\begin{equation}
	g^{00}_{(1)} = 0.
\end{equation}
Thus the requirement that the \KGe have solutions which are formal power series of the form \eqref{eq:WKB_ansatz} imposes restrictions on the components of the metric. In the following, we will freely use the vanishing of $g^{00}_{(1)}$.

Using these results, the positive frequency \KGe for our WKB-like solutions is equivalent to an equation for $\psi$ and further, inserting the expansion $\psi = \sum_{k=0}^\infty c^{-k} a_k$, to a collective equation for the $a_k$. These equations are given as \eqref{eq:WKB_KG_psi} and \eqref{eq:WKB_KG} in the appendix.

Using \eqref{eq:WKB_KG}, we can obtain equations for the $a_k$, order by order, which can then be combined into a Schrödinger equation for $\psi$: At order $c^0$, we have
\begin{equation}
	0 = \left(-\frac{m^2}{\hbar^2} g^{00}_{(2)} - \frac{\I m}{\hbar} (\partial_i g^{0i}_{(1)}) - \frac{2\I m}{\hbar}g^{0i}_{(1)} \partial_i + \frac{2\I m}{\hbar} \partial_t + \Delta\right) a_0,
\end{equation}
i.e. the Schrödinger equation
\begin{equation} \label{eq:Schroedinger_WKB_exp_0}
	\I\hbar \partial_t a_0 = \left(-\frac{\hbar^2}{2m} \Delta + \frac{\I\hbar}{2} (\partial_i g^{0i}_{(1)}) + \I\hbar g^{0i}_{(1)} \partial_i + \frac{m}{2} g^{00}_{(2)}\right) a_0.
\end{equation}
By the relation $\psi = a_0 + \Or(c^{-1})$, this also gives a Schrödinger equation for $\psi$ in 0$^\mathrm{th}$ order in $c^{-1}$.

At order $c^{-1}$, \eqref{eq:WKB_KG} yields the following Schrödinger-like equation for $a_1$ with correction terms involving $a_0$:
\begin{eqnarray}\fl \eqalign{ \label{eq:Schroedinger_WKB_exp_1}
	\I\hbar \partial_t a_1 &= \left(-\frac{\hbar^2}{2m} \Delta + \frac{\I\hbar}{2} (\partial_i g^{0i}_{(1)}) + \I\hbar g^{0i}_{(1)} \partial_i + \frac{m}{2} g^{00}_{(2)}\right) a_1\\
		&\quad +\bigg(-\frac{\I \hbar}{4} g^{0i}_{(1)} [\partial_i \tr(\eta g^{-1}_{(1)})] + \frac{\I \hbar}{4} [\partial_t \tr(\eta g^{-1}_{(1)})] + \frac{\hbar^2}{4m} [\partial_i \tr(\eta g^{-1}_{(1)})] \partial_i\\
			&\qquad - \frac{\hbar^2}{2m} (\partial_i g^{ij}_{(1)}) \partial_j - \frac{\hbar^2}{2m} g^{ij}_{(1)} \partial_i\partial_j + \frac{m}{2} g^{00}_{(3)} + \frac{\I \hbar}{2} (\partial_i g^{0i}_{(2)}) + \I \hbar g^{0i}_{(2)} \partial_i\bigg) a_0
}\end{eqnarray}
Using $\psi = a_0 + c^{-1} a_1 + \Or(c^{-2})$, we can combine \eqref{eq:Schroedinger_WKB_exp_1} with \eqref{eq:Schroedinger_WKB_exp_0} into a Schrödinger equation for $\psi$ up to order $c^{-1}$:
\begin{eqnarray}\fl \eqalign{ \label{eq:Schroedinger_WKB}
	\I\hbar \partial_t \psi &= \Bigg[-\frac{\hbar^2}{2m} \Delta + \frac{\I\hbar}{2} (\partial_i g^{0i}_{(1)}) + \I\hbar g^{0i}_{(1)} \partial_i + \frac{m}{2} g^{00}_{(2)} + c^{-1} \bigg(-\frac{\I \hbar}{4} g^{0i}_{(1)} [\partial_i \tr(\eta g^{-1}_{(1)})]\\
		&\qquad + \frac{\I \hbar}{4} [\partial_t \tr(\eta g^{-1}_{(1)})] + \frac{\hbar^2}{4m} [\partial_i \tr(\eta g^{-1}_{(1)})] \partial_i - \frac{\hbar^2}{2m} (\partial_i g^{ij}_{(1)}) \partial_j - \frac{\hbar^2}{2m} g^{ij}_{(1)} \partial_i\partial_j \\
		&\qquad + \frac{m}{2} g^{00}_{(3)} + \frac{\I \hbar}{2} (\partial_i g^{0i}_{(2)}) + \I \hbar g^{0i}_{(2)} \partial_i\bigg) + \Or(c^{-2})\Bigg] \psi =: H \psi.
}\end{eqnarray}

Continuing this process of evaluating \eqref{eq:WKB_KG}, we can, in principle, get Schrödinger equations for $\psi$ to arbitrary order in $c^{-1}$, i.e. obtain the Hamiltonian in the Schrödinger form of the positive frequency \KGe to arbitrary order in $c^{-1}$.

However, when considering higher orders, a difficulty arises: The Schrödinger-like equations for $a_k$ begin to involve time derivatives of the lower order functions $a_l$, so we have to re-use the derived equations for the $a_l$ in order to get a true Schrödinger equation for $\psi$ (with a purely `spatial' Hamiltonian, i.e. not involving any time derivatives), i.e. the process becomes recursive.
As far as concrete calculations up to some finite order are concerned, this is merely a computational obstacle; but for a general analysis of the expansion method this poses a bigger problem, since no general closed form can be easily obtained.
This motivated the study of the \KGe as a quadratic equation for the time derivative operator, leading to the method described in section \ref{sec:mom_exp}.

%%%%%%%%%%%%%%%%%%%%%%%%%%%%%%%%%%%%%%%%%%%%%%%%%%%%%%%%%%%
\subsection{Transformation to `flat' scalar product and comparison with canonical quantisation} \label{subsec:WKB_trafo_flat}

To transform the Hamiltonian obtained in \eqref{eq:Schroedinger_WKB} to the `flat' scalar product in order to compare it to the result from canonical quantisation, we note that for two positive frequency solutions $\Psi_\mathrm{KG} = \exp(-\I mc^2 t/\hbar) \psi$ and $\Phi_\mathrm{KG} = \exp(-\I mc^2 t/\hbar) \varphi$, the Klein--Gordon inner product is given by
\begin{eqnarray}\fl \eqalign{ \label{eq:KG_ip}
	\langle\Psi_\mathrm{KG},\Phi_\mathrm{KG}\rangle_\mathrm{KG} &= \I\hbar c \int g^{0\nu} [(\partial_\nu \overline{\Psi_\mathrm{KG}}) \Phi_\mathrm{KG} - \overline{\Psi_\mathrm{KG}} (\partial_\nu \Phi_\mathrm{KG})] \frac{1}{\sqrt{-g^{00}}} \sqrt{^{(3)}g} \D^3x\\
	&= \int \Bigg(\sqrt{-g^{00}} \left[2mc^2 \overline{\psi}\varphi + \overline{(H \psi)}\varphi + \overline\psi (H \varphi)\right]
		\\&\qquad + \I\hbar c \frac{g^{0i}}{\sqrt{-g^{00}}} \left[\overline{(\partial_i \psi)}\varphi - \overline\psi(\partial_i \varphi)\right] \Bigg)\sqrt{^{(3)}g} \D^3x,
}\end{eqnarray}
where we used our adapted coordinates and chose $\Sigma = \{t = \mathrm{const.}\}$ in the general form \eqref{eq:KG_ip_general} of the Klein--Gordon inner product.

Using $\sqrt{-g^{00}} = 1 + \Or(c^{-2})$, $g^{0i} (-g^{00})^{-1/2} = \Or(c^{-1})$ and $H = \Or(c^{0})$, we get
\begin{equation}
	\frac{1}{2mc^2} \langle \Psi_\mathrm{KG}, \Phi_\mathrm{KG} \rangle_\mathrm{KG} = \int [\overline{\psi}\varphi + \Or(c^{-2})]\sqrt{^{(3)}g} \D^3x.
\end{equation}
For this to equal the `flat' scalar product $\int \overline{\psi_f} \varphi_f \D^3 x$, we see that the `flat wavefunction' has to have the form $\psi_f = {^{(3)}g^{1/4}} \psi + \Or(c^{-2})$ and therefore evolves according to the Schrödinger equation $\I\hbar \partial_t \psi_f = H_f \psi_f$ with the `flat Hamiltonian'
\begin{equation}
	H_f = \I\hbar(\partial_t {^{(3)}g^{1/4}}){^{(3)}g^{-1/4}} + {^{(3)}g^{1/4}} H ({^{(3)}g^{-1/4}} \cdot) + \Or(c^{-2}).
\end{equation}

Using\footnote
	{The determinant satisfies $g^{-1} = -1 - c^{-1} \tr(\eta g^{-1}_{(1)}) + \Or(c^{-2})$. Using the well-known identity ${^{(3)}g} = g^{00} g$ for a $3+1$ decomposed metric, this gives ${^{(3)}g} = g^{00} g = [-1 + \Or(c^{-2})] [-1 + c^{-1} \tr(\eta g^{-1}_{(1)}) + \Or(c^{-2})] = 1 - c^{-1} \tr(\eta g^{-1}_{(1)}) + \Or(c^{-2})$.}
${^{(3)}g^{1/4}} = 1 - c^{-1}\frac{1}{4} \tr(\eta g^{-1}_{(1)}) + \Or(c^{-2})$ and noting that conjugation with a multiplication operator leaves multiplication operators invariant, we obtain
\begin{eqnarray}\fl \eqalign{ \label{eq:Schroedinger_WKB_flat}
	H_f &= -\I\hbar c^{-1} \frac{1}{4} [\partial_t \tr(\eta g^{-1}_{(1)})] + H - c^{-1}\frac{\hbar^2}{8m} [\Delta, \tr(\eta g^{-1}_{(1)})]\\
		&\qquad + c^{-1} \frac{\I\hbar}{4} g^{0i}_{(1)} [\partial_i, \tr(\eta g^{-1}_{(1)})] + \Or(c^{-2})\\
	&= -\I\hbar c^{-1} \frac{1}{4} [\partial_t \tr(\eta g^{-1}_{(1)})] + H - c^{-1}\frac{\hbar^2}{8m} \left([\Delta \tr(\eta g^{-1}_{(1)})] + 2[\partial_i \tr(\eta g^{-1}_{(1)})] \partial_i\right) \\
		&\qquad + c^{-1} \frac{\I\hbar}{4} g^{0i}_{(1)} [\partial_i \tr(\eta g^{-1}_{(1)})] + \Or(c^{-2})\\
	&= -\frac{\hbar^2}{2m} \Delta + \frac{\I\hbar}{2} (\partial_i g^{0i}_{(1)}) + \I\hbar g^{0i}_{(1)} \partial_i + \frac{m}{2} g^{00}_{(2)} + c^{-1} \bigg(-\frac{\hbar^2}{2m} (\partial_i g^{ij}_{(1)}) \partial_j - \frac{\hbar^2}{2m} g^{ij}_{(1)} \partial_i\partial_j\\
		&\qquad + \frac{m}{2} g^{00}_{(3)} + \frac{\I \hbar}{2} (\partial_i g^{0i}_{(2)}) + \I \hbar g^{0i}_{(2)} \partial_i - \frac{\hbar^2}{8m} [\Delta \tr(\eta g^{-1}_{(1)})]\bigg) + \Or(c^{-2})\\
	&= -\frac{\hbar^2}{2m} \Delta - \frac{1}{2} \left\{g^{0i}_{(1)}, -\I\hbar\partial_i\right\} + \frac{m}{2} g^{00}_{(2)} + c^{-1} \bigg(\frac{1}{2m} (-\I\hbar) \partial_i \Big(g^{ij}_{(1)} (-\I\hbar)\partial_j \cdot \Big)\\
		&\qquad + \frac{m}{2} g^{00}_{(3)} - \frac{1}{2} \left\{g^{0i}_{(2)}, -\I\hbar\partial_i\right\} - \frac{\hbar^2}{8m} [\Delta \tr(\eta g^{-1}_{(1)})]\bigg) + \Or(c^{-2}),
}\end{eqnarray}
where $\{A,B\} = AB + BA$ denotes the anticommutator. This is the Hamiltonian appearing in the `flat' Schrödinger form of the positive frequency \KGe up to order $c^{-1}$, obtained by the WKB-like approximation in a general metric.

For comparison of this result with the canonical quantisation scheme, we have to subtract the rest energy $mc^2$ from the classical Hamiltonian of equation \eqref{eq:class_Ham}, corresponding to the phase factor separated off the Klein--Gordon field, and expand it in $c^{-1}$, yielding
\begin{eqnarray} \fl \eqalign{
	H_\mathrm{class} &= \frac{1}{\sqrt{-g^{00}}} c \left[m^2 c^2 + \left(g^{ij} - \frac{1}{g^{00}} g^{0i} g^{0j}\right) p_i p_j\right]^{1/2} - mc^2 + \frac{c}{g^{00}} g^{0i} p_i\\
	&= \frac{m}{2} g^{00}_{(2)} + \frac{p_i p_i}{2m} - g^{0i}_{(1)} p_i + c^{-1} \left(\frac{m}{2} g^{00}_{(3)} + g^{ij}_{(1)} \frac{p_i p_j}{2m} - g^{0i}_{(2)} p_i\right) + \Or(c^{-2}).
}\end{eqnarray}
Comparing this with \eqref{eq:Schroedinger_WKB_flat}, we see that by `canonical quantisation' of this classical Hamiltonian using the rule `$p_i \to -\I\partial_i$', we can reproduce, using a special ordering scheme, all terms appearing in the WKB expansion, apart from $-\frac{\hbar^2}{8mc} [\Delta \tr(\eta g^{-1}_{(1)})]$. For this last term to arise by naive canonical quantisation, consisting only of symmetrising according to some ordering scheme and replacing momenta by operators, in the classical Hamiltonian there would have to be a term proportional to $\frac{p_i p_i}{mc} \tr(\eta g^{-1}_{(1)}) = \frac{p_i p_i}{mc} g^{jj}_{(1)}$, which is not the case.

As the most simple non-trivial example, for the `Newtonian' metric with line element
\begin{equation}
	\D s^2 = -\left(1+2\frac{\phi}{c^2}\right)c^2\D t^2 + \delta_{ij} \D x^i \D x^j + \Or(c^{-2}),
\end{equation}
the inverse metric has components
\begin{equation}
	(g^{\mu\nu}) = \left(\begin{array}{cc}-1 + 2 \frac{\phi}{c^2} + \Or(c^{-4}) & \Or(c^{-3})\\ \Or(c^{-3}) & \mathds{1} + \Or(c^{-2}) \end{array}\right),
\end{equation}
leading to the quantum Hamiltonian $H = -\frac{\hbar^2}{2m}\Delta + m \phi + \Or(c^{-2})$ in both schemes, i.e. just the standard Hamiltonian with Newtonian potential.

The occurrence of an extra term in a geometrically motivated quantum theory which one cannot arrive at by naive canonical quantisation is reminiscent of the occurrence of a `quantum-mechanical potential' term in the Hamiltonian found by DeWitt in his 1952 treatment of quantum motion in a curved space \cite{dewitt52}:
By demanding the (free part of the) Hamiltonian to be given by $H^\mathrm{DeWitt} = -\frac{\hbar^2}{2m} {^{(3)}\hspace{-.2em}\Delta_\mathrm{LB}}$ in terms of the spatial Laplace--Beltrami operator $^{(3)}\hspace{-.2em}\Delta_\mathrm{LB}$, it turns out to have the form $H^\mathrm{DeWitt} = \frac{1}{2m}\hat p_i g^{ij} \hat p_j + \hbar^2 Q$ of a sum of a naively canonically quantised kinetic term and the quantum-mechanical potential\footnote
	{Using the form $-\hbar^2 \, {^{(3)}\hspace{-.2em} \Delta_\mathrm{LB}} = {^{(3)}g^{-1/4}} \, \hat p_i \, {^{(3)}g^{1/2}} \, {^{(3)}g^{ij}} \, \hat p_j \, {^{(3)}g^{-1/4}}$ of the Laplace--Beltrami operator in terms of the momentum operator \eqref{eq:canon_mom_op_induced}, it can be expressed as $-\hbar^2 \, {^{(3)}\hspace{-.2em} \Delta_\mathrm{LB}} = \hat p_i g^{ij} p_j - {^{(3)}g^{-1/4}} [\hat p_i, {^{(3)}g^{ij}} [\hat p_j, {^{(3)}g^{1/4}}]]$, giving the above expression for the quantum-mechanical potential.}
$\hbar^2 Q = \frac{\hbar^2}{2m} {^{(3)}g^{-1/4}} \partial_i ({^{(3)}g^{ij}} \partial_j {^{(3)}g^{1/4}})$.

In fact, for our metric \eqref{eq:exp_metric}, in lowest order in $c^{-1}$ the quantum-mechanical potential is given by $\hbar^2 Q = -\frac{\hbar^2}{8mc}\Delta g^{ii}_{(1)} + \Or(c^{-2}) = -\frac{\hbar^2}{8mc}[\Delta \tr(\eta g^{-1}_{(1)})]  + \Or(c^{-2})$, thus reproducing the additional term arising in the WKB method. This connection of the WKB-like expansion to the three-dimensional `spatial' geometry seems interesting, but to further investigate it, a more geometric, coordinate-free formulation of the expansion, possibly starting with a choice of $3+1$ decomposition of spacetime, ought to be used.

One could argue that DeWitt's Hamiltonian \emph{can} be arrived at by canonical quantisation in some sense, since the Laplace--Beltrami operator can be written as $- \hbar^2 \, {^{(3)}\hspace{-.2em} \Delta_\mathrm{LB}} = {^{(3)}g^{-1/4}} \, \hat p_i \, {^{(3)}g^{1/2}} \, {^{(3)}g^{ij}} \, \hat p_j \, {^{(3)}g^{-1/4}}$ in terms of the momentum operator \eqref{eq:canon_mom_op_induced} corresponding to the `geometric' scalar product \eqref{eq:canon_scalar_prod_induced} which was used by DeWitt. However, such a `clever rewriting' of the Newtonian kinetic term in the classical Hamiltonian as $\frac{1}{2m} g^{ij} p_i p_j = \frac{1}{2m} {^{(3)}g^{-1/4}} \, p_i \, {^{(3)}g^{1/2}} \, {^{(3)}g^{ij}} \, p_j \, {^{(3)}g^{-1/4}}$ before replacing momenta by operators involves more than just choosing some symmetrised operator ordering and thus is not part of what we called `canonical quantisation' above.

%%%%%%%%%%%%%%%%%%%%%%%%%%%%%%%%%%%%%%%%%%%%%%%%%%%%%%%%%%%
\subsection{The Eddington--Robertson PPN metric as an explicit example}

The Eddington--Robertson parametrised post-Newtonian metric is given by the line element
\begin{equation} \fl
	\D s^2 = -\left(1 + 2\frac{\phi}{c^2} + 2 \beta \frac{\phi^2}{c^4}\right)c^2\D t^2 + \left(1 - 2 \gamma \frac{\phi}{c^2}\right)\delta_{ij} \D x^i \D x^j + \Or(c^{-3})
\end{equation}
with the Eddington-Robertson parameters $\beta,\gamma$. For the case of general relativity, both these parameters take the value $1$. The components of the metric can be read off from the expression for the line element to be
\begin{equation}
	(g_{\mu\nu}) =
	\left(\begin{array}{cc}
		-1 - 2 \frac{\phi}{c^2} - 2 \beta \frac{\phi^2}{c^4} + \Or(c^{-5}) & \Or(c^{-4})\\
		\Or(c^{-4}) & \left(1 - 2 \gamma \frac{\phi}{c^2}\right)\mathds{1} + \Or(c^{-3})
	\end{array}\right),
\end{equation}
leading to the inverse metric having components
\begin{equation}
	\fl (g^{\mu\nu}) =
	\left(\begin{array}{cc}
	-1 + 2 \frac{\phi}{c^2} + (2\beta-4) \frac{\phi^2}{c^4} + \Or(c^{-5}) & \Or(c^{-4})\\
	\Or(c^{-4}) & \left(1 + 2 \gamma \frac{\phi}{c^2}\right)\mathds{1} + \Or(c^{-3})
	\end{array}\right).
\end{equation}

The equations arising for $a_0, a_1$ from \eqref{eq:WKB_KG} at orders $c^0, c^{-1}$ are thus simply the Schrödinger equations
\begin{equation}
	\I\hbar \partial_t a_i = \left(-\frac{\hbar^2}{2m} \Delta + m\phi\right) a_i, \quad i = 0,1;
\end{equation}
at order $c^{-2}$, we get
\begin{eqnarray} \fl \eqalign{
	0 &= \bigg[- \frac{\I m}{\hbar} (\partial_t g^{00}_{(2)}) - \frac{2\I m}{\hbar} g^{00}_{(2)} \partial_t - \partial_t^2 + \frac{\I m}{2\hbar} \bigg( -[\partial_t \tr(\eta g^{-1}_{(2)})] + \frac{1}{2} [\partial_t \tr(\eta \underbrace{g^{-1}_{(2,2)}}_{\makebox[0pt]{\scriptsize $= g^{-1}_{(1)} \eta g^{-1}_{(1)} = 0$}} )] \bigg)\\
			&\qquad + \frac{1}{2} \left( -[\partial_i \tr(\eta g^{-1}_{(2)})] \partial_i + \frac{1}{2} [\partial_i \tr(\eta g^{-1}_{(2,2)})] \partial_i \right) + (\partial_i g^{ij}_{(2)}) \partial_j + g^{ij}_{(2)} \partial_i\partial_j - \frac{m^2}{\hbar^2} g^{00}_{(4)} \bigg] a_0\\
		&\quad + \left(-\frac{m^2}{\hbar^2} g^{00}_{(2)} + \frac{2\I m}{\hbar} \partial_t + \Delta\right) a_2,\\
	&= \bigg(- \frac{4\I m}{\hbar} \phi \partial_t - \partial_t^2 - \frac{\I m}{\hbar} (3\gamma + 1) (\partial_t \phi) - (\gamma - 1) (\partial_i \phi) \partial_i \\
			&\qquad + 2\gamma \phi \Delta - \frac{m^2}{\hbar^2} (2\beta - 4) \phi^2 \bigg) a_0\\
		&\quad + \left(-\frac{2 m^2}{\hbar^2} \phi + \frac{2\I m}{\hbar} \partial_t + \Delta\right) a_2,
}\end{eqnarray}
or equivalently the Schrödinger-like equation
\begin{eqnarray} \eqalign{
	\I\hbar \partial_t a_2 &= \left(-\frac{\hbar^2}{2m} \Delta + m\phi\right) a_2 + \bigg(2\I\hbar \phi \partial_t + \frac{\hbar^2}{2m} \partial_t^2 + \frac{\I\hbar}{2} (3\gamma + 1) (\partial_t \phi)\\
		&\qquad + \frac{\hbar^2}{2m} (\gamma - 1) (\partial_i \phi) \partial_i - \frac{\hbar^2}{m} \gamma \phi \Delta + \frac{m}{2} (2\beta - 4) \phi^2 \bigg) a_0.
}\end{eqnarray}
Using the Schrödinger equation for $a_0$, we have
\begin{eqnarray} \fl \eqalign{
	\frac{\hbar^2}{2m} \partial_t^2 a_0 &= - \frac{\I\hbar}{2m} \partial_t \left(-\frac{\hbar^2}{2m} \Delta + m\phi\right) a_0 = -\frac{\I\hbar}{2} (\partial_t \phi) a_0 - \frac{1}{2m} \left(-\frac{\hbar^2}{2m} \Delta + m\phi\right) \I\hbar \partial_t a_0\\
	&= -\frac{\I\hbar}{2} (\partial_t \phi) a_0 - \frac{1}{2m} \left(-\frac{\hbar^2}{2m} \Delta + m\phi\right)^2 a_0\\
	&= -\frac{\I\hbar}{2} (\partial_t \phi) a_0 -\frac{\hbar^4}{8m^3} \Delta \Delta a_0 + \frac{\hbar^2}{4m} \Delta (\phi a_0) + \frac{\hbar^2}{4m} \phi \Delta a_0 - \frac{m}{2} \phi^2 a_0\\
	&= -\frac{\I\hbar}{2} (\partial_t \phi) a_0 -\frac{\hbar^4}{8m^3} \Delta \Delta a_0 + \frac{\hbar^2}{4m} (\Delta\phi) a_0 + \frac{\hbar^2}{2m} (\partial_i \phi) \partial_i a_0 + \frac{\hbar^2}{2m} \phi \Delta a_0 - \frac{m}{2} \phi^2 a_0,
}\end{eqnarray}
and thus the equation for $a_2$ becomes
\begin{eqnarray} \eqalign{
	\I\hbar \partial_t a_2 &= \left(-\frac{\hbar^2}{2m} \Delta + m\phi\right) a_2
	+ \bigg(-\frac{\hbar^4}{8m^3} \Delta \Delta + \frac{\hbar^2}{4m} (\Delta\phi) + \frac{3 \I\hbar}{2} \gamma (\partial_t \phi)\\&\qquad + \frac{\hbar^2}{2m} \gamma (\partial_i \phi) \partial_i - \frac{\hbar^2}{2m} (2\gamma + 1) \phi \Delta + \frac{m}{2} (2\beta - 1) \phi^2 \bigg) a_0.
}\end{eqnarray}
Combining the equations for $a_0, a_1, a_2$, the Hamiltonian in the Schrödinger equation $\I\hbar \partial_t \psi = H \psi$ for the `wavefunction' (i.e. phase-shifted Klein--Gordon field) $\psi$ thus reads
\begin{eqnarray} \eqalign{
	H &= -\frac{\hbar^2}{2m} \Delta + m\phi
	+ \frac{1}{c^2}\bigg(-\frac{\hbar^4}{8m^3} \Delta \Delta + \frac{\hbar^2}{4m} (\Delta\phi) + \frac{3 \I\hbar}{2} \gamma (\partial_t \phi)\\&\qquad + \frac{\hbar^2}{2m} \gamma (\partial_i \phi) \partial_i - \frac{\hbar^2}{2m} (2\gamma + 1) \phi \Delta + \frac{m}{2} (2\beta - 1) \phi^2 \bigg) + \Or(c^{-3}),
}\end{eqnarray}
reproducing, up to notational differences and the fact that we did not consider coupling to an electromagnetic field, the result of Lämmerzahl \cite[eq. (8)]{laemmerzahl95}.

To transform to the flat scalar product, we note that in our metric and using this Hamiltonian, the Klein--Gordon inner product \eqref{eq:KG_ip} is given by
\begin{equation}
	\frac{1}{2mc^2} \langle\Psi_\mathrm{KG},\Phi_\mathrm{KG}\rangle_\mathrm{KG} = \int \left(\overline{\psi}\varphi - \frac{\hbar^2}{2m^2c^2} \overline{\psi} \Delta\varphi + \Or(c^{-3}) \right) \sqrt{^{(3)}g} \D^3x.
\end{equation}
For this to equal the flat scalar product $\int \overline{\psi_f} \varphi_f \D^3x$, the flat wavefunction has to have the form $\psi_f = \left(1 - \frac{\hbar^2}{2m^2c^2}\Delta\right)^{1/2} {^{(3)}g^{1/4}} \psi + \Or(c^{-3})$ (note that $\frac{1}{c^2} \Delta$ commutes with $^{(3)}g$ up to higher-order terms), resulting in the flat Hamiltonian
\begin{eqnarray} \fl \eqalign{
	H_f &= \I\hbar(\partial_t {^{(3)}g^{1/4}}){^{(3)}g^{-1/4}}\\
		&\quad+ \left(1 - \frac{\hbar^2}{2m^2c^2}\Delta\right)^{1/2} {^{(3)}g^{1/4}} H \, {^{(3)}g^{-1/4}} \left(1 - \frac{\hbar^2}{2m^2c^2}\Delta\right)^{-1/2} + \Or(c^{-3}).
}\end{eqnarray}
Using ${^{(3)}g^{1/4} = 1 - \frac{3}{2} \gamma \frac{\phi}{c^2} + \Or(c^{-3})}$ and $\left(1 - \frac{\hbar^2}{2m^2c^2}\Delta\right)^{1/2} = 1 - \frac{\hbar^2}{4m^2c^2}\Delta + \Or(c^{-4})$, this yields
\begin{eqnarray} \fl \eqalign{
	H_f &= -\I\hbar\left(\partial_t \frac{3}{2} \gamma \frac{\phi}{c^2} \right) + H + \left[-\frac{3}{2}\gamma\frac{\phi}{c^2}, -\frac{\hbar^2}{2m} \Delta\right] + \left[- \frac{\hbar^2}{4m^2c^2}\Delta, m\phi\right] + \Or(c^{-3})\\
	&= -\frac{3\I\hbar}{2c^2} \gamma (\partial_t \phi) + H - \frac{\hbar^2}{4mc^2} (3\gamma + 1) [\Delta, \phi] + \Or(c^{-3})\\
	&= -\frac{3\I\hbar}{2c^2} \gamma (\partial_t \phi) + H - \frac{\hbar^2}{4mc^2} (3\gamma + 1) ((\Delta \phi) + 2 (\partial_i \phi) \partial_i) + \Or(c^{-3})\\
	&= -\frac{\hbar^2}{2m} \Delta + m\phi
	+ \frac{1}{c^2}\bigg(-\frac{\hbar^4}{8m^3} \Delta \Delta - \frac{3\hbar^2}{4m} \gamma (\Delta\phi) \\&\qquad - \frac{\hbar^2}{2m} (2\gamma + 1) (\partial_i \phi) \partial_i - \frac{\hbar^2}{2m} (2\gamma + 1) \phi \Delta + \frac{m}{2} (2\beta - 1) \phi^2 \bigg) + \Or(c^{-3}),
}\end{eqnarray}
reproducing the flat Hamiltonian of Lämmerzahl \cite[eq. (16)]{laemmerzahl95}.

In comparison, the classical Hamiltonian (minus the rest energy) expands to
\begin{eqnarray} \fl \eqalign{
	H_\mathrm{class} &= \frac{1}{\sqrt{-g^{00}}} c \left[m^2 c^2 + \left(g^{ij} - \frac{1}{g^{00}} g^{0i} g^{0j}\right) p_i p_j\right]^{1/2} - mc^2 + \frac{c}{g^{00}} g^{0i} p_i\\
	&= \frac{p_i p_i}{2m} + m\phi + c^{-2} \left(-\frac{(p_i p_i)^2}{8m^3} + \frac{m \phi^2}{2}(2\beta - 1) + \frac{\phi}{2m}(2\gamma + 1) p_i p_i\right) + \Or(c^{-3}).
}\end{eqnarray}
By canonical quantisation of this, we cannot reproduce the Hamiltonian obtained from the WKB expansion in the case of a general $\gamma$, but just for some special choices of $\gamma$, depending on the ordering scheme: For example, in the anticommutator ordering scheme, we would quantise the classical function $\phi p_i p_i$ as $\frac{1}{2} \{-\hbar^2\Delta, \phi\} = -\frac{\hbar^2}{2} (\Delta\phi) - \hbar^2 (\partial_i\phi) \partial_i - \hbar^2 \phi \Delta$, reproducing the WKB Hamiltonian in the case of $\gamma = 1$; but when quantising it as $-\hbar^2 \partial_i (\phi \partial_i \cdot) = -\hbar^2 (\partial_i \phi) \partial_i -\hbar^2 \phi \Delta$, this would lead to agreement with the WKB Hamiltonian for $\gamma = 0$.

%%%%%%%%%%%%%%%%%%%%%%%%%%%%%%%%%%%%%%%%%%%%%%%%%%%%%%%%%%%
\section{General comparison of the two methods by momentum expansion} \label{sec:mom_exp}

We will now describe a method by which general statements about similarities and differences between the two approaches explained above can be made in the case of stationary spacetimes, without any `non-relativistic' expansion in $c^{-1}$. Instead, we consider `potential' terms and terms linear, quadratic, \ldots\ in momentum, i.e. we perform a (formal) expansion in momenta.

%%%%%%%%%%%%%%%%%%%%%%%%%%%%%%%%%%%%%%%%%%%%%%%%%%%%%%%%%%%
\subsection{The \KGe as a quadratic equation for the Hamiltonian}

Assume a \emph{stationary} spacetime and work in adapted coordinates $(x^0 = ct, x^i)$, i.e. coordinates such that $\partial_t$ is (a constant multiple of) the stationarity Killing field. In particular, we have $\partial_t g_{\mu\nu} = 0$.
The coordinate expression for the d'Alembert operator on functions is thus
\begin{eqnarray} \eqalign{
	\Box f &= \frac{1}{\sqrt{-g}} \partial_\mu(\sqrt{-g} g^{\mu\nu} \partial_\nu f)\\
	&= \frac{1}{\sqrt{-g}} (\partial_\mu\sqrt{-g}) g^{\mu\nu} \partial_\nu f + (\partial_\mu g^{\mu\nu}) \partial_\nu f + g^{\mu\nu} \partial_\mu\partial_\nu f\\
	&= \frac{1}{2g} (\partial_i g) g^{i\nu} \partial_\nu f + (\partial_i g^{i\nu}) \partial_\nu f + g^{\mu\nu} \partial_\mu\partial_\nu f.
}\end{eqnarray}
Hence, the minimally coupled \KGe reads
\begin{eqnarray} \eqalign{
	0 &= \left(\Box - \frac{m^2c^2}{\hbar^2}\right)\Psi\\
	&= \frac{1}{c} \frac{1}{2g} (\partial_i g) g^{0i} \partial_t \Psi + \frac{1}{2g} (\partial_i g) g^{ij} \partial_j \Psi + \frac{1}{c} (\partial_i g^{0i}) \partial_t \Psi + (\partial_i g^{ij}) \partial_j \Psi\\
	&\qquad + \frac{1}{c^2} g^{00} \partial_t^2 \Psi + \frac{2}{c} g^{0i} \partial_i\partial_t \Psi + g^{ij} \partial_i\partial_j \Psi - \frac{m^2c^2}{\hbar^2}\Psi.
}\end{eqnarray}
This means that the space of solutions of the \KGe is the kernel of $\mathcal P(\I\hbar \partial_t)$, where for an operator $A$ acting on the functions on the spacetime, $\mathcal P(A)$ is the following operator:
\begin{eqnarray} \eqalign{
	\mathcal P(A) = &-\frac{\I}{\hbar c} \frac{1}{2g} (\partial_i g) g^{0i} A + \frac{1}{2g} (\partial_i g) g^{ij} \partial_j - \frac{\I}{\hbar c} (\partial_i g^{0i}) A + (\partial_i g^{ij}) \partial_j\\
	&- \frac{1}{\hbar^2 c^2} g^{00} A^2 - \frac{2\I}{\hbar c} g^{0i} \partial_i\circ A + g^{ij} \partial_i\partial_j - \frac{m^2c^2}{\hbar^2}
}\end{eqnarray}

Thus, wanting to write the \KGe in the form of a Schrödinger equation $\I\hbar\partial_t \Psi = H \Psi$ -- and thus restricting to the solutions of the \KGe for which this is possible -- we see that this can be achieved by demanding the Hamiltonian $H$ to be a solution of the quadratic operator equation
\begin{equation} \label{eq:KG_operator_expression}
	0 = \mathcal P(H)
\end{equation}
and be composed only of spatial derivative operators and coefficients of the metric, not involving any time derivatives: Stationarity of the metric then implies $[\partial_t, H] = 0$, such that the Schrödinger equation yields $(\I\hbar\partial_t)^2 \Psi = \I\hbar\partial_t H \Psi = H \I\hbar\partial_t \Psi = H^2 \Psi$, leading to $\mathcal P(\I\hbar\partial_t) \Psi = \mathcal P(H) \Psi = 0$ by \eqref{eq:KG_operator_expression}; i.e. every solution of the Schrödinger equation is also a solution of the \KGe.

In the following, we will solve equation \eqref{eq:KG_operator_expression} by expanding $H$ as a formal power series in spatial derivative operators, i.e. momentum operators. The two possible solutions we will obtain for $H$ correspond to positive and negative frequency solutions of the \KGe, respectively.

%%%%%%%%%%%%%%%%%%%%%%%%%%%%%%%%%%%%%%%%%%%%%%%%%%%%%%%%%%%
\subsection{Momentum expansion and first-order solution}

We expand $H$ as $H = H_{(0)} + H_{(1)} + \Or(\partial_i^2)$, where $H_{(k)}$ includes all terms involving $k$ spatial derivative operators. Using this notation, the lowest order term of \eqref{eq:KG_operator_expression}, involving no spatial derivatives, reads
\begin{equation}
	0 = -\frac{1}{\hbar^2 c^2} g^{00} H_{(0)}^2 - \frac{m^2c^2}{\hbar^2},
\end{equation}
giving
\begin{equation} \label{eq:KG_operator_H0}
	H_{(0)} = \frac{mc^2}{\sqrt{-g^{00}}}
\end{equation}
where we choose the positive square root since we are interested in positive frequency solutions of the \KGe.

At order $\partial_i^1$, equation \eqref{eq:KG_operator_expression} gives
\begin{eqnarray} \eqalign{
	0 = &-\frac{\I}{\hbar c} \frac{1}{2g} (\partial_i g) g^{0i} H_{(0)} - \frac{\I}{\hbar c} (\partial_i g^{0i}) H_{(0)}\\
	&- \frac{1}{\hbar^2 c^2} g^{00} (2 H_{(0)} H_{(1)} + [H_{(1)}, H_{(0)}]) - \frac{2\I}{\hbar c} g^{0i} \partial_i\circ H_{(0)}.
}\end{eqnarray}
Writing $H_{(1)} = H_{(1,M)} + H_{(N,C)}^i \partial_i$ where $H_{(1,M)}$ is a multiplication operator (involving one spatial differentiation of some function) and $H_{(N,C)}^i$ are coefficient functions not involving any differentiations, we have $[H_{(1)}, H_{(0)}] = [H_{(1,C)}^i \partial_i, H_{(0)}] = H_{(1,C)}^i (\partial_i H_{(0)})$. Thus, the equation reads
\begin{eqnarray} \eqalign{
	0 = &-\frac{\I}{\hbar c} \frac{1}{2g} (\partial_i g) g^{0i} H_{(0)} - \frac{\I}{\hbar c} (\partial_i g^{0i}) H_{(0)} - \frac{2g^{00}}{\hbar^2 c^2} H_{(0)} H_{(1)}\\
	&- \frac{g^{00}}{\hbar^2 c^2} H_{(1,C)}^i (\partial_i H_{(0)}) - \frac{2\I}{\hbar c} g^{0i} (\partial_i H_{(0)}) - \frac{2\I}{\hbar c} g^{0i} H_{(0)} \partial_i.
}\end{eqnarray}
The right-hand side now has two different components: A multiplication operator and an operator differentiating the function it acts upon. These components have to vanish independently. The `differentiating part' is $0 = - \frac{2g^{00}}{\hbar^2 c^2} H_{(0)} H_{(1,C)}^i \partial_i - \frac{2\I}{\hbar c} g^{0i} H_{(0)} \partial_i$, equivalently
\begin{equation} \label{eq:KG_operator_H1C}
	H_{(1,C)}^i = - \I \hbar c \frac{g^{0i}}{g^{00}}.
\end{equation}
Thus, the multiplication operator part is
\begin{equation}
	\fl 0 = -\frac{\I}{\hbar c} \frac{1}{2g} (\partial_i g) g^{0i} H_{(0)} - \frac{\I}{\hbar c} (\partial_i g^{0i}) H_{(0)} - \frac{2g^{00}}{\hbar^2 c^2} H_{(0)} H_{(1,M)} - \frac{\I}{\hbar c} g^{0i} (\partial_i H_{(0)}),
\end{equation}
giving
\begin{equation} \label{eq:KG_operator_H1M}
	H_{(1,M)} = -\frac{\I\hbar c}{4g^{00} g} (\partial_i g) g^{0i} - \frac{\I\hbar c}{2g^{00}} (\partial_i g^{0i}) - \frac{\I\hbar c}{2g^{00}} g^{0i} \frac{1}{H_{(0)}} (\partial_i H_{(0)}).
\end{equation}
Since $\frac{1}{H_{(0)}} (\partial_i H_{(0)}) = \sqrt{-g^{00}} \partial_i \frac{1}{\sqrt{-g^{00}}} = \frac{g^{00}}{2} \partial_i \frac{1}{g^{00}}$, \eqref{eq:KG_operator_H0}, \eqref{eq:KG_operator_H1C} and \eqref{eq:KG_operator_H1M} together yield the result
\begin{equation}\fl \label{eq:KG_operator_H_complete}
	H = \frac{mc^2}{\sqrt{-g^{00}}} - \frac{\I\hbar c}{4g^{00} g} (\partial_i g) g^{0i} - \frac{\I\hbar c}{2g^{00}} (\partial_i g^{0i}) - \frac{\I\hbar c}{4} g^{0i} \left(\partial_i \frac{1}{g^{00}}\right) - \I \hbar c \frac{g^{0i}}{g^{00}} \partial_i + \Or(\partial_i^2)
\end{equation}
for the Hamiltonian in the Schrödinger form
\begin{equation}
	\I\hbar \partial_t \Psi = H \Psi
\end{equation}
of the positive frequency \KGe, at first order in momenta.

%%%%%%%%%%%%%%%%%%%%%%%%%%%%%%%%%%%%%%%%%%%%%%%%%%%%%%%%%%%
\subsection{Transformation to `flat' scalar product and comparison with canonical quantisation}

To transform this Hamiltonian to the `flat' scalar product, we note that for two positive frequency solutions $\Psi$ and $\Phi$, the Klein--Gordon inner product is given by
\begin{eqnarray} \eqalign{
	\langle\Psi,\Phi\rangle_\mathrm{KG} &= \I\hbar c \int g^{0\nu} [(\partial_\nu \overline\Psi)\Phi - \overline\Psi(\partial_\nu \Phi)] \frac{1}{\sqrt{-g^{00}}} \sqrt{^{(3)}g} \D^3x\\
	&= \int \Bigg(\sqrt{-g^{00}} \left[\overline{(H \Psi)}\Phi + \overline\Psi(H \Phi)\right]\\
		&\qquad+ \I\hbar c \frac{g^{0i}}{\sqrt{-g^{00}}} \left[\overline{(\partial_i \Psi)}\Phi - \overline\Psi(\partial_i \Phi)\right] \Bigg)\sqrt{^{(3)}g} \D^3x\\
	\eqref{eq:KG_operator_H_complete} \quad &= \int 2mc^2 \overline\Psi \Phi\sqrt{^{(3)}g} \D^3x + \Or(\partial_i^2).
}\end{eqnarray}
For this to equal the `flat' scalar product $\int \overline{\Psi_f} \Phi_f \D^3 x$, we see that the `flat wavefunction' has to have the form $\Psi_f = \sqrt{2mc^2} {^{(3)}g^{1/4}} \Psi + \Or(\partial_i^2)$ and therefore evolves according to the Schrödinger equation $\I\hbar \partial_t \Psi_f = H_f \Psi_f$ with the `flat Hamiltonian'
\begin{equation}
	H_f = {^{(3)}g^{1/4}} H ({^{(3)}g^{-1/4}} \cdot) + \Or(\partial_i^2).
\end{equation}
For calculating $H_f$ from $H$, we note that conjugating with a multiplication operator leaves multiplication operators invariant and that
\begin{eqnarray} \eqalign{
	{^{(3)}g^{1/4}} \partial_i({^{(3)}g^{-1/4}} \cdot) &= \partial_i - \frac{1}{4} [\partial_i \ln({^{(3)}g})]\\
	&= \partial_i - \frac{1}{4} [\partial_i \ln(g^{00}g)]\\
	&= \partial_i - \frac{1}{4} \frac{1}{g} (\partial_i g) - \frac{1}{4} \frac{1}{g^{00}} (\partial_i g^{00}),
}\end{eqnarray}
yielding the final result
\begin{eqnarray} \eqalign{
	H_f &= \frac{mc^2}{\sqrt{-g^{00}}} - \frac{\I\hbar c}{4g^{00} g} (\partial_i g) g^{0i} - \frac{\I\hbar c}{2g^{00}} (\partial_i g^{0i}) - \frac{\I\hbar c}{4} g^{0i} \left(\partial_i \frac{1}{g^{00}}\right)\\
		&\quad - \I \hbar c \frac{g^{0i}}{g^{00}} \left(\partial_i - \frac{1}{4} \frac{1}{g} (\partial_i g) - \frac{1}{4} \frac{1}{g^{00}} (\partial_i g^{00})\right) + \Or(\partial_i^2)\\
	&= \frac{mc^2}{\sqrt{-g^{00}}} - \frac{\I\hbar c}{2} (\partial_i \frac{g^{0i}}{g^{00}}) - \I \hbar c \frac{g^{0i}}{g^{00}} \partial_i + \Or(\partial_i^2)\\
	&= \frac{mc^2}{\sqrt{-g^{00}}} + c \frac{1}{2} \left\{ \frac{g^{0i}}{g^{00}}, -\I\hbar \partial_i \right\} + \Or(\partial_i^2).
}\end{eqnarray}

Looking at the momentum expansion of the classical Hamiltonian
\begin{eqnarray} \eqalign{
	H_\mathrm{class} &= \frac{1}{\sqrt{-g^{00}}} c \left[m^2 c^2 + \left(g^{ij} - \frac{1}{g^{00}} g^{0i} g^{0j}\right) p_i p_j\right]^{1/2} + \frac{c}{g^{00}} g^{0i} p_i\\
	&= \frac{mc^2}{\sqrt{-g^{00}}} + \frac{c}{g^{00}} g^{0i} p_i + \Or(p_i^2),
}\end{eqnarray}
we see that `canonical quantisation' of this Hamiltonian will lead to the same `potential term' as did the Klein--Gordon equation, regardless of the adopted ordering scheme, and to the same term linear in momentum if we use any ordering scheme leading to `anticommutator quantisation' for linear terms.

%%%%%%%%%%%%%%%%%%%%%%%%%%%%%%%%%%%%%%%%%%%%%%%%%%%%%%%%%%%
\section{Conclusion}

We have shown how to derive a Schrödinger equation with post-Newtonian correction terms describing a single quantum particle in a general curved background spacetime by means of a WKB-like formal expansion of the minimally coupled \KGe. We extended this method to account for, in principle, terms of arbitrary orders in $c^{-1}$, although it gets recursive at higher orders, making it computationally more difficult to handle than methods based on formal quantisation of the classical description of the particle. Nevertheless, we believe this scheme to be better suited for concrete predictions, since it is more firmly based on first principles and also more systematic than \emph{ad hoc} canonical quantisation procedures as employed widely in the literature. For example, no operator ordering ambiguities arise; instead, the WKB method can be seen as \emph{predicting} the ordering.

Comparing the Klein--Gordon expansion method to canonical quantisation, we have found that in the case of a general metric, even at lowest post-Newtonian order, the two procedures lead to slightly different quantum Hamiltonians, \emph{independent of ordering ambiguities}\footnote
	{At least if only simple symmetrising procedures are allowed for as ordering schemes in canonical quantisation, see the discussion at the end of section \ref{subsec:WKB_trafo_flat}.}.
For the concrete case of the metric of the Eddington--Robertson PPN test theory, the Hamiltonians obtained from the two methods differ in a term including the Eddington--Robertson parameter $\gamma$, depending on the ordering scheme employed in canonical quantisation. Thus for the interpretation of tests of general relativity with quantum systems, the method used to derive the quantum Hamiltonian plays a decisive rôle.

For the case of stationary background metrics, without employing any `non-relativistic' expansion, we showed that up to linear order in spatial momenta, the Hamiltonians obtained from canonical quantisation of a free particle and from the Klein--Gordon equation agree. In particular, this means that the lowest-order coupling to the `gravitomagnetic' field $g^{0i}$ is independent of the gravity--quantum matter coupling method.

It should be clear that the classical Klein--Gordon equation, as well as any expansion based on it, meets its limits as soon as effects of particle creation and annihilation become relevant. Obviously, these cannot be accounted for in a single-particle description as proposed here. One may indeed wonder whether and how this single-particle scheme can be generalised to a system of a fixed number of many interacting particles. This is not as straightforward as one might think at first. It is well known that there exist severe obstructions against formulating a relativistic theory of many interacting particles; see, e.g., chapter 21 of \cite{sudarshan16}, where the `no-interaction-theorems' in Poincaré invariant Hamiltonian mechanics are discussed.

Concerning the applicability of the WKB-like method for concrete calculations, it is also an interesting question if and how the transformation of the Hamiltonian from the Klein--Gordon inner product to an $\mathrm L^2$-scalar product -- be it flat or with respect to the induced measure $\sqrt{^{(3)}g} \D^3x$ -- can be implemented more systematically, not relying on direct calculations with the already-computed Hamiltonian.

Another important issue to address is the coordinate-dependence of the methods developed in this paper: It would be desirable to have at hand a coordinate-free formulation of the expansion/coupling schemes with clear geometric interpretation, thus possibly shedding some light on the connection of the WKB-like expansion to the spatial geometry of the Cauchy surfaces and respecting the geometric nature of gravity.

%%%%%%%%%%%%%%%%%%%%%%%%%%%%%%%%%%%%%%%%%%%%%%%%%%%%%%%%%%%
\section*{Acknowledgements}
The authors would like to thank two anonymous referees for valuable comments on the manuscript.

This work was supported by the Deutsche Forschungsgemeinschaft through the Collaborative Research Centre 1227 (DQ-mat), project B08.

%%%%%%%%%%%%%%%%%%%%%%%%%%%%%%%%%%%%%%%%%%%%%%%%%%%%%%%%%%%
\appendix

%%%%%%%%%%%%%%%%%%%%%%%%%%%%%%%%%%%%%%%%%%%%%%%%%%%%%%%%%%%
\section{Calculation of the classical Hamiltonian of a free particle} \label{app:details_class_Ham}

We will now give a full exposition of the calculation of the classical Hamiltonian of a free particle in a curved spacetime in $3+1$ decomposition.

In $3+1$ decomposition, spacetime is foliated into 3-dimensional spacelike Cauchy surfaces that are labelled by a `foliation parameter' $t$. We employ adapted coordinates $(x^0 = ct, x^i)$ where $x^i$ are coordinates on these Cauchy surfaces.
This gives a decomposition of the spacetime metric as $g_{ij} = h_{ij}, g_{0i} = h_{ij} \beta^j =: \beta_i, g_{00} = -\alpha^2 + h_{ij} \beta^i \beta^j$ where $h$ is the induced metric on the Cauchy surfaces, $\beta$ is the shift vector field and $\alpha$ is the lapse function.
Geometrically speaking, lapse and shift arise from decomposing the `time evolution' vector field\footnote
	{Denoting the embeddings defining the foliation as $\mathcal E_t\colon \Sigma \to M, \; t \in \mathbb R$ where $\Sigma$ is the abstract Cauchy surface, the time evolution vector field is given as the derivation \[\left.\frac{\partial}{\partial t}\right|_{\mathcal E_s(q)} f := \left.\frac{\D}{\D s'} f(\mathcal E_{s'}(q))\right|_{s' = s}\] for $q\in \Sigma$, $s\in\mathbb R$ and $f \in C^\infty(M)$; i.e. this vector field is independent of the choice of coordinates and depends just on the foliation, even if it was expressed above as a coordinate vector field. \cite[eq. (17.43)]{giulini14}}
$\partial_0 = c^{-1} \partial/\partial t$ into its components tangential and normal to the Cauchy surfaces as
\begin{equation}
	\frac{1}{c} \frac{\partial}{\partial t} = \alpha n + \beta,
\end{equation}
where $n$ is the future-directed unit normal to the Cauchy surfaces and $\beta$ is the tangential component \cite[eq. (17.44)]{giulini14}.

Parametrising the worldline of a free particle by $t$, its Lagrangian (compare the classical action \eqref{eq:class_action}) in these coordinates is
\begin{equation}\fl \label{eq:class_Lagr}
	L = -mc \sqrt{-g_{\mu\nu} \dot x^\mu \dot x^\nu} = -mc \left(\alpha^2 c^2 - h_{ij} \beta^i \beta^j c^2 - 2 c h_{ij} \dot x^i \beta^j - h_{ij} \dot x^i \dot x^j\right)^{1/2},
\end{equation}
where a dot denotes differentiation with respect to $t$.
From this, we compute the momentum $p_i$ conjugate to $x^i$ to be
\begin{equation}
	p_i = \frac{\partial L}{\partial \dot x^i} = \frac{mc}{(\ldots)^{1/2}}(c \beta_i + h_{ij} \dot x^j).
\end{equation}
Contracting with the inverse $h^{ij}$ of $h_{ij}$, we obtain
\begin{equation} \label{eq:dotx}
	\dot x^i = \frac{(\ldots)^{1/2}}{mc} h^{ij}p_j - c \beta^i.
\end{equation}

To fully express the velocity $\dot x^i$ in terms of the momentum $p_i$, we have to express $(\ldots)^{1/2} = \left(\alpha^2 c^2 - h(\dot{\vec x} + c \vec \beta, \dot{\vec x} + c \vec \beta)\right)^{1/2}$ in terms of $p_i$. Using \eqref{eq:dotx}, we have
\begin{eqnarray} \eqalign{
	h(\dot{\vec x} + c \vec \beta, \dot{\vec x} + c \vec \beta) &= \frac{(\ldots)}{m^2c^2} h^{ij} p_i p_j\\
	&= \frac{\alpha^2 c^2 - h(\dot{\vec x} + c \vec\beta, \dot{\vec x} + c \vec\beta)}{m^2 c^2} h^{ij} p_i p_j.
}\end{eqnarray}
Writing $h^{-1}(p,p) := h^{ij} p_i p_j$, this is equivalent to
\begin{eqnarray} \eqalign{
	h(\dot{\vec x} + c \vec\beta, \dot{\vec x} + c \vec\beta) &= \frac{\alpha^2 h^{-1}(p,p)}{m^2} \frac{1}{1 + h^{-1}(p,p) / (m^2 c^2)}\\
	&= \frac{c^2 \alpha^2 h^{-1}(p,p)}{m^2c^2 + h^{-1}(p,p)}
}\end{eqnarray}
Using this, we get
\begin{equation}\fl \label{eq:sqrt_expression}
	(\ldots)^{1/2} = \left(\alpha^2 c^2 - \frac{c^2 \alpha^2 h^{-1}(p,p)}{m^2 c^2 + h^{-1}(p,p)}\right)^{1/2} = \frac{m c^2 \alpha}{\left[m^2 c^2 + h^{-1}(p,p)\right]^{1/2}}.
\end{equation}

Inserting \eqref{eq:sqrt_expression} into \eqref{eq:dotx}, we can express the velocities in terms of the canonical momenta as
\begin{equation} \label{eq:dotx_momentum}
	\dot x^i = \frac{\alpha c}{\left[m^2 c^2 + h^{-1}(p,p)\right]^{1/2}} h^{ij} p_j - c \beta^i.
\end{equation}
Using \eqref{eq:sqrt_expression} and \eqref{eq:dotx_momentum}, the Hamiltonian corresponding to the Lagrangian \eqref{eq:class_Lagr} can be computed to be
\begin{eqnarray} \eqalign{
	H &= p_i \dot x^i - L\\
	&= \frac{\alpha c}{\left[m^2 c^2 + h^{-1}(p,p)\right]^{1/2}} h^{-1}(p,p) - c \beta^i p_i + \frac{m^2 c^3 \alpha}{\left[m^2 c^2 + h^{-1}(p,p)\right]^{1/2}}\\
	&= \alpha c \left[m^2 c^2 + h^{-1}(p,p)\right]^{1/2} - c \beta^i p_i.
}\end{eqnarray}

Rewriting this in terms of the components of the spacetime metric using the relations $g^{00} = -\alpha^{-2}, g^{0i} = \alpha^{-2} \beta^i, g^{ij} = h^{ij} - \alpha^{-2} \beta^i \beta^j$, the Hamiltonian reads
\begin{equation}
	H = \frac{1}{\sqrt{-g^{00}}} m c^2 \left[1 + \left(g^{ij} - \frac{1}{g^{00}} g^{0i} g^{0j}\right) \frac{p_i p_j}{m^2 c^2}\right]^{1/2} + \frac{c}{g^{00}} g^{0i} p_i.
\end{equation}

%%%%%%%%%%%%%%%%%%%%%%%%%%%%%%%%%%%%%%%%%%%%%%%%%%%%%%%%%%%
\section{Details of the WKB expansion} \label{app:details_WKB}

We will now explicitly compute the three terms in the expression \eqref{eq:box_op} for the d'Alembert operator. Inserting the expansion \eqref{eq:exp_metric} of the components of the inverse metric and using $x^0 = ct$, the third term is
\begin{eqnarray}\fl \eqalign{ \label{eq:box_expanded_term3}
	g^{\mu\nu} \partial_\mu\partial_\nu &= -c^{-2}\partial_t^2 + \Delta + \sum_{k=1}^\infty c^{-k} g^{00}_{(k)} c^{-2} \partial_t^2 + \sum_{k=1}^\infty c^{-k} 2 g^{0i}_{(k)} c^{-1} \partial_t\partial_i + \sum_{k=1}^\infty c^{-k} g^{ij}_{(k)} \partial_i\partial_j\\
	&= -c^{-2}\partial_t^2 + \Delta + \sum_{k=3}^\infty c^{-k} g^{00}_{(k-2)} \partial_t^2 + \sum_{k=2}^\infty c^{-k} 2 g^{0i}_{(k-1)} \partial_t\partial_i + \sum_{k=1}^\infty c^{-k} g^{ij}_{(k)} \partial_i\partial_j.
}\end{eqnarray}
Similarly, we get
\begin{eqnarray}\eqalign{ \label{eq:box_expanded_term2}
	(\partial_\mu g^{\mu\nu})\partial_\nu &= \sum_{k=3}^\infty c^{-k} (\partial_t g^{00}_{(k-2)}) \partial_t + \sum_{k=2}^\infty c^{-k} (\partial_t g^{0i}_{(k-1)}) \partial_i \\&\qquad + \sum_{k=2}^\infty c^{-k} (\partial_i g^{0i}_{(k-1)}) \partial_t + \sum_{k=1}^\infty c^{-k} (\partial_i g^{ij}_{(k)}) \partial_j
}\end{eqnarray}
for the second term.

Using the form
\begin{equation}
	g^{\mu\nu} = \left[ \left(\mathds{1} + \sum_{k=1}^\infty c^{-k} g^{-1}_{(k)} \eta\right) \eta^{-1} \right]^{\mu\nu}
\end{equation}
of the expanded inverse metric, we can use a formal Neumann series to invert the power series and write the coefficients of the metric as
\begin{equation}
	g_{\mu\nu} = \left\{ \eta \left[\mathds{1} + \sum_{n=1}^\infty \left(-\sum_{k=1}^\infty c^{-k} g^{-1}_{(k)} \eta\right)^n \right] \right\}_{\mu\nu}.
\end{equation}
Iterating the Cauchy product formula, we have
\begin{equation}
	\left(-\sum_{k=1}^\infty c^{-k} g^{-1}_{(k)} \eta\right)^n = (-1)^n \sum_{k=1}^\infty c^{-k} \sum_{i_1+\cdots+i_n = k \atop 1 \le i_1,\ldots,i_n \le k} g^{-1}_{(i_1)}\eta \cdots g^{-1}_{(i_n)}\eta.
\end{equation}
Using this and the notation
\begin{equation}
	g^{-1}_{(k,n)} := \sum_{i_1+\cdots+i_n = k \atop 1 \le i_1,\ldots,i_n \le k} g^{-1}_{(i_1)}\eta g^{-1}_{(i_2)}\eta \cdots g^{-1}_{(i_n)}
\end{equation}
introduced in the main text, we can write the metric as
\begin{equation}
	g_{\mu\nu} = \eta_{\mu\nu} + \sum_{k=1}^\infty c^{-k} \sum_{n=1}^\infty (-1)^n (\eta g^{-1}_{(k,n)} \eta)_{\mu\nu}.
\end{equation}
Thus, using the Cauchy product formula again, we get
\begin{eqnarray} \fl \eqalign{
	g_{\rho\sigma}\partial_\mu g^{\rho\sigma} &= \left(\eta_{\rho\sigma} + \sum_{k=1}^\infty c^{-k} \sum_{n=1}^\infty (-1)^n (\eta g^{-1}_{(k,n)} \eta)_{\rho\sigma}\right) \sum_{m=1}^\infty c^{-m} \partial_\mu g^{\rho\sigma}_{(m)}\\
	&= \sum_{k=1}^\infty c^{-k} \partial_\mu \tr(\eta g^{-1}_{(k)}) + \sum_{k=2}^\infty c^{-k} \sum_{l + m = k} \sum_{n=1}^\infty (-1)^n (\eta g^{-1}_{(l,n)} \eta)_{\rho\sigma} \partial_\mu g^{\rho\sigma}_{(m)}.
}\end{eqnarray}
Using
\begin{eqnarray} \eqalign{
	\sum_{l + m = k} (\eta g^{-1}_{(l,n)} \eta)_{\rho\sigma} \partial_\mu g^{\rho\sigma}_{(m)} &= \sum_{l + m = k} \sum_{i_1+\cdots+i_n = l} (\eta g^{-1}_{(i_1)} \cdots g^{-1}_{(i_n)} \eta)_{\rho\sigma} \partial_\mu g^{\rho\sigma}_{(m)}\\
	&= \sum_{i_1+\cdots+i_n + m = k} \tr(\eta g^{-1}_{(i_1)} \cdots g^{-1}_{(i_n)} \eta \partial_\mu g^{-1}_{(m)})\\
	&= \frac{1}{n+1} \partial_\mu \sum_{i_1+\cdots+i_n + m = k} \tr(\eta g^{-1}_{(i_1)} \cdots g^{-1}_{(i_n)} \eta g^{-1}_{(m)})\\
	&= \frac{1}{n+1} \partial_\mu \tr(\eta g^{-1}_{(k,n+1)}),
}\end{eqnarray}
we can rewrite this as
\begin{eqnarray} \eqalign{
	g_{\rho\sigma}\partial_\mu g^{\rho\sigma} &= \sum_{k=1}^\infty c^{-k} \partial_\mu \tr(\eta g^{-1}_{(k)}) + \sum_{k=2}^\infty c^{-k} \sum_{n=2}^\infty (-1)^{n-1} \frac{1}{n} \partial_\mu \tr(\eta g^{-1}_{(k,n)})\\
	&= \sum_{k=1}^\infty c^{-k} \partial_\mu \tr(\eta g^{-1}_{(k)}) + \sum_{k=1}^\infty c^{-k} \sum_{n=2}^\infty (-1)^{n-1} \frac{1}{n} \partial_\mu \tr(\eta g^{-1}_{(k,n)})\\
	&= \sum_{k=1}^\infty c^{-k} \sum_{n=1}^\infty (-1)^{n-1} \frac{1}{n} \partial_\mu \tr(\eta g^{-1}_{(k,n)})
}\end{eqnarray}
since $g^{-1}_{(k,n)} = 0$ for $n > k$ and $g^{-1}_{(k,1)} = g^{-1}_{(k)}$.
Thus, we finally have the expansion
\begin{eqnarray}\fl \eqalign{ \label{eq:box_expanded_term1}
	&\frac{1}{\sqrt{-g}} (\partial_\mu\sqrt{-g}) g^{\mu\nu} \partial_\nu f\\
	&\qquad= -\frac{1}{2} (g_{\rho\sigma} \partial_\mu g^{\rho\sigma}) g^{\mu\nu} \partial_\nu f\\
	&\qquad= \frac{1}{2} \sum_{k=1}^\infty c^{-k} \sum_{n=1}^\infty (-1)^n \frac{1}{n} [\partial_\mu \tr(\eta g^{-1}_{(k,n)})] \left(\eta^{\mu\nu} + \sum_{m=1}^\infty c^{-m} g^{\mu\nu}_{(m)} \right) \partial_\nu f\\
	&\qquad= \frac{1}{2} \sum_{k=1}^\infty c^{-k} \sum_{n=1}^\infty (-1)^n \frac{1}{n} [\partial_\mu \tr(\eta g^{-1}_{(k,n)})] \eta^{\mu\nu} \partial_\nu f \\&\qquad\qquad + \frac{1}{2} \sum_{k=2}^\infty c^{-k} \sum_{l+m = k} \sum_{n=1}^\infty (-1)^n \frac{1}{n} [\partial_\mu \tr(\eta g^{-1}_{(l,n)})] g^{\mu\nu}_{(m)} \partial_\nu f
}\end{eqnarray}
for the first term in the d'Alembert operator \eqref{eq:box_op}.

Inserting \eqref{eq:box_expanded_term3},\eqref{eq:box_expanded_term2} and \eqref{eq:box_expanded_term1} into \eqref{eq:box_op} and sorting the sums by order of $c^{-1}$, we get the expansion \eqref{eq:box_expanded} for the d'Alembert operator.

The fully expanded positive frequency \KGe as an equation for the function $\psi$ in the Klein--Gordon field $\Psi_\mathrm{KG} = \exp\left(\I m c^2 t/\hbar\right) \psi$ is as follows:
\begin{eqnarray}\fl \eqalign{ \label{eq:WKB_KG_psi}
	0 &= \sum_{k=5}^\infty c^{-k} \frac{1}{2} \sum_{l+m = k-2} \sum_{n=1}^\infty (-1)^n \frac{1}{n} g^{00}_{(m)} [\partial_t \tr(\eta g^{-1}_{(l,n)})] \partial_t \psi\\
	&\qquad + \sum_{k=4}^\infty c^{-k} (\partial_t g^{00}_{(k-2)}) \partial_t \psi + \sum_{k=4}^\infty c^{-k} g^{00}_{(k-2)} \partial_t^2 \psi\\
	&\qquad - \sum_{k=3}^\infty c^{-k} \frac{\I m}{2\hbar} \sum_{l+m = k} \sum_{n=1}^\infty (-1)^n \frac{1}{n} g^{00}_{(m)} [\partial_t \tr(\eta g^{-1}_{(l,n)})] \psi\\
	&\qquad + \sum_{k=3}^\infty c^{-k} \frac{1}{2} \sum_{l+m = k-1} \sum_{n=1}^\infty (-1)^n \frac{1}{n} g^{0i}_{(m)} \left( [\partial_t \tr(\eta g^{-1}_{(l,n)})] \partial_i \psi + [\partial_i \tr(\eta g^{-1}_{(l,n)})] \partial_t \psi\right)\\
	&\qquad - \sum_{k=3}^\infty c^{-k} \frac{1}{2} \sum_{n=1}^\infty (-1)^n \frac{1}{n} [\partial_t \tr(\eta g^{-1}_{(k-2,n)})] \partial_t \psi\\
	&\qquad - \sum_{k=2}^\infty c^{-k} \frac{\I m}{\hbar} (\partial_t g^{00}_{(k)}) \psi - \sum_{k=2}^\infty c^{-k} \frac{2\I m}{\hbar} g^{00}_{(k)} \partial_t \psi\\
	&\qquad + \sum_{k=2}^\infty c^{-k} \frac{1}{2} \sum_{l+m = k} \sum_{n=1}^\infty (-1)^n \frac{1}{n} g^{ij}_{(m)} [\partial_i \tr(\eta g^{-1}_{(l,n)})] \partial_j \psi\\
	&\qquad + \sum_{k=2}^\infty c^{-k} \left((\partial_t g^{0i}_{(k-1)}) \partial_i \psi + (\partial_i g^{0i}_{(k-1)}) \partial_t \psi\right) + \sum_{k=2}^\infty c^{-k} 2 g^{0i}_{(k-1)} \partial_t\partial_i \psi - c^{-2} \partial_t^2 \psi\\
	&\qquad - \sum_{k=1}^\infty c^{-k} \frac{\I m}{2\hbar} \sum_{l+m = k+1} \sum_{n=1}^\infty (-1)^n \frac{1}{n} g^{0i}_{(m)} [\partial_i \tr(\eta g^{-1}_{(l,n)})] \psi\\
	&\qquad + \sum_{k=1}^\infty c^{-k} \frac{\I m}{2\hbar} \sum_{n=1}^\infty (-1)^n \frac{1}{n} [\partial_t \tr(\eta g^{-1}_{(k,n)})] \psi\\
	&\qquad + \sum_{k=1}^\infty c^{-k} \frac{1}{2} \sum_{n=1}^\infty (-1)^n \frac{1}{n} [\partial_i \tr(\eta g^{-1}_{(k,n)})] \partial_i \psi + \sum_{k=1}^\infty c^{-k} (\partial_i g^{ij}_{(k)}) \partial_j \psi\\
	&\qquad + \sum_{k=1}^\infty c^{-k} g^{ij}_{(k)} \partial_i\partial_j \psi\\
	&\qquad - \sum_{k=0}^\infty c^{-k} \frac{m^2}{\hbar^2} g^{00}_{(k+2)} \psi - \sum_{k=0}^\infty c^{-k} \frac{\I m}{\hbar} (\partial_i g^{0i}_{(k+1)}) \psi - \sum_{k=0}^\infty c^{-k} \frac{2\I m}{\hbar} g^{0i}_{(k+1)} \partial_i \psi\\
	&\qquad + \frac{2\I m}{\hbar} \partial_t \psi + \Delta \psi.
}\end{eqnarray}

Inserting the expansion $\psi = \sum_{k=0}^\infty c^{-k} a_k$ and using the Cauchy product formula, this is equivalent to
\begin{eqnarray}\fl \eqalign{ \label{eq:WKB_KG}
	0 &= \sum_{k=5}^\infty c^{-k} \frac{1}{2} \sum_{l+m + \tilde k = k-2} \sum_{n=1}^\infty (-1)^n \frac{1}{n} g^{00}_{(m)} [\partial_t \tr(\eta g^{-1}_{(l,n)})] \partial_t a_{\tilde k}\\
	&\qquad + \sum_{k=4}^\infty c^{-k} \sum_{l + \tilde k = k-2} (\partial_t g^{00}_{(l)}) \partial_t a_{\tilde k} + \sum_{k=4}^\infty c^{-k} \sum_{l + \tilde k = k-2} g^{00}_{(l)} \partial_t^2 a_{\tilde k}\\
	&\qquad - \sum_{k=3}^\infty c^{-k} \frac{\I m}{2\hbar} \sum_{l+m + \tilde k = k} \sum_{n=1}^\infty (-1)^n \frac{1}{n} g^{00}_{(m)} [\partial_t \tr(\eta g^{-1}_{(l,n)})] a_{\tilde k}\\
	&\qquad + \sum_{k=3}^\infty c^{-k} \frac{1}{2} \sum_{l+m + \tilde k = k-1} \sum_{n=1}^\infty (-1)^n \frac{1}{n} g^{0i}_{(m)} \left([\partial_t \tr(\eta g^{-1}_{(l,n)})] \partial_i a_{\tilde k} + [\partial_i \tr(\eta g^{-1}_{(l,n)})] \partial_t a_{\tilde k}\right)\\
	&\qquad - \sum_{k=3}^\infty c^{-k} \frac{1}{2} \sum_{l + \tilde k = k-2} \sum_{n=1}^\infty (-1)^n \frac{1}{n} [\partial_t \tr(\eta g^{-1}_{(l,n)})] \partial_t a_{\tilde k}\\
	&\qquad - \sum_{k=2}^\infty c^{-k} \frac{\I m}{\hbar} \sum_{l + \tilde k = k} (\partial_t g^{00}_{(l)}) a_{\tilde k} - \sum_{k=2}^\infty c^{-k} \frac{2\I m}{\hbar} \sum_{l + \tilde k = k} g^{00}_{(l)} \partial_t a_{\tilde k}\\
	&\qquad + \sum_{k=2}^\infty c^{-k} \frac{1}{2} \sum_{l+m + \tilde k = k} \sum_{n=1}^\infty (-1)^n \frac{1}{n} g^{ij}_{(m)} [\partial_i \tr(\eta g^{-1}_{(l,n)})] \partial_j a_{\tilde k}\\
	&\qquad + \sum_{k=2}^\infty c^{-k} \sum_{l + \tilde k = k-1} \left((\partial_t g^{0i}_{(l)}) \partial_i a_{\tilde k} + (\partial_i g^{0i}_{(l)}) \partial_t a_{\tilde k}\right) + \sum_{k=2}^\infty c^{-k} 2 \sum_{l + \tilde k = k-1} g^{0i}_{(l)} \partial_t\partial_i a_{\tilde k}\\
	&\qquad - \sum_{k=2}^\infty c^{-k} \partial_t^2 a_{k-2} - \sum_{k=1}^\infty c^{-k} \frac{\I m}{2\hbar} \sum_{l+m + \tilde k = k+1} \sum_{n=1}^\infty (-1)^n \frac{1}{n} g^{0i}_{(m)} [\partial_i \tr(\eta g^{-1}_{(l,n)})] a_{\tilde k}\\
	&\qquad + \sum_{k=1}^\infty c^{-k} \frac{\I m}{2\hbar} \sum_{l + \tilde k = k} \sum_{n=1}^\infty (-1)^n \frac{1}{n} [\partial_t \tr(\eta g^{-1}_{(l,n)})] a_{\tilde k}\\
	&\qquad + \sum_{k=1}^\infty c^{-k} \frac{1}{2} \sum_{l + \tilde k = k} \sum_{n=1}^\infty (-1)^n \frac{1}{n} [\partial_i \tr(\eta g^{-1}_{(l,n)})] \partial_i a_{\tilde k} + \sum_{k=1}^\infty c^{-k} \sum_{l + \tilde k = k} (\partial_i g^{ij}_{(l)}) \partial_j a_{\tilde k}\\
	&\qquad + \sum_{k=1}^\infty c^{-k} \sum_{l + \tilde k = k} g^{ij}_{(l)} \partial_i\partial_j a_{\tilde k} - \sum_{k=0}^\infty c^{-k} \frac{m^2}{\hbar^2} \sum_{l + \tilde k = k+2} g^{00}_{(l)} a_{\tilde k}\\
	&\qquad - \sum_{k=0}^\infty c^{-k} \frac{\I m}{\hbar} \sum_{l + \tilde k = k+1} (\partial_i g^{0i}_{(l)}) a_{\tilde k} - \sum_{k=0}^\infty c^{-k} \frac{2\I m}{\hbar} \sum_{l + \tilde k = k+1} g^{0i}_{(l)} \partial_i a_{\tilde k}\\
	&\qquad + \sum_{k=0}^\infty c^{-k} \frac{2\I m}{\hbar} \partial_t a_k + \sum_{k=0}^\infty c^{-k} \Delta a_k,
}\end{eqnarray}
where in sums like $\sum_{l+m+\tilde k = k}$, $l$ and $m$ are $\ge 1$ as before, but $\tilde k$ is $\ge 0$.

\section*{References}
\bibliographystyle{iopart-num}
\bibliography{references}

\end{document}